\documentclass{article}

\usepackage[margin=1in]{geometry}
\usepackage{graphicx} 
\usepackage{cite}
\usepackage{hyperref}
\usepackage{amsmath, multirow}
\usepackage{amssymb}
\usepackage{slashed}
\usepackage{comment}
\usepackage[export]{adjustbox}

\usepackage[rightcaption]{sidecap}

\newcommand{\be}{\begin{equation}}
\newcommand{\ee}{\end{equation}}
\newcommand{\bea}{\begin{eqnarray}}
\newcommand{\eea}{\end{eqnarray}}

\title{General Signals for Charged Lepton Flavor Violating Decays}
\author{Spencer Chang and Thomas Driscoll\\
Department of Physics and Institute for Fundamental Science\\ 
University of Oregon, Eugene, Oregon 97403}

\renewcommand{\arraystretch}{1.25}

\begin{document}

\maketitle

\begin{abstract}
We explore the most general phenomenology of charged lepton flavor violating (CLFV) decays of muon and tau leptons to the three body final states $(\bar{e}ee, \bar{\mu}\mu\mu, \bar{e}\mu\mu, \bar{\mu}\mu e,\bar{\mu}ee, \bar{e}e\mu)$. By constructing a complete basis of operators at each dimension, we derive the most general amplitudes for these decay processes.  By considering constraints from unitarity and LEP, we show that operators of mass dimension 6 and 7 are the most likely to be observed in next generation experiments. Focusing on these dimensions, we compute the results of unpolarized (spin-averaged) decays parameterized in terms of the invariant masses of the daughter particles. We also compute the differential decay rates for polarized decays, in anticipation of the experimental search Mu3e, which expects to have a muon beam with $\sim 90\%$ polarization, and the Chiral Belle proposal, which aims to have a $70\%$ polarized electron beam.  To determine the extent to which the operators may be distinguished experimentally, we plot the differential distributions for each operator, showing that they leave only a few possible degenerate explanations. Through a statistical analysis, we estimate the number of events needed to break the degeneracies using the angular information.  These results are adapted to treat $\ell\to \ell' \nu \bar{\nu}$, where the angular distribution of the outgoing charged lepton has enhanced distinguishing power.  With many Standard Model extensions predicting these CLFV decays, these results will better enable upcoming searches to identify and/or constrain physics beyond the Standard Model. 

\end{abstract}
\newpage
\section{Introduction}

A myriad of well motivated SM extensions require that charged lepton flavor violating (CLFV) interactions exist, from new physics at energy scales inaccessible to current colliders. These include  various models for neutrino masses \cite{Konetschny:1977bn, Magg:1980ut, Cheng:1980qt, Foot:1988aq, Zee:1980ai, Petcov:1976ff, Babu:1988ki, Ma:2006km}, as well as certain supersymmetric and grand unified theories \cite{Gabbiani:1988rb,Ilakovac:1994kj, Barbieri:1995tw, Hisano:1995cp, Hisano:1998fj, Deppisch:2004fa, Bambhaniya:2015ipg}. While the new particles of these theories may be out of reach of colliders, their low energy effects may be observed through the interactions they mediate virtually.    In the Standard Model, rates for charged lepton flavor-violating decays, such as $\mu \rightarrow e\gamma$ and $\mu \rightarrow eee$, receive negligibly small contributions due to nonzero neutrino masses  \cite{Petcov:1976ff, Marciano:1977wx, Lee:1977qz, Lee:1977tib, Hernandez-Tome:2018fbq, Blackstone:2019njl}.\footnote{The caveat is if there are extremely light Majorana neutrinos, see \cite{Calibbi:2017uvl}.}  
As these predictions are dramatically out of reach of experimental probes, any observation of CLFV would be a definitive sign of new physics. 

To look for CLFV effects in the most general way, it is useful to use an effective field theory (EFT) approach.  Existing EFT analyses have predominantly been performed in the Standard Model EFT (SMEFT) framework \cite{Fernandez-Martinez:2024bxg, Ardu:2023yyw, Ardu:2024bua, Cirigliano:2021img, Grzadkowski:2010es}.  However, SMEFT cannot describe general scenarios where there are either new sources of electroweak symmetry breaking or heavy particles, whose mass predominantly comes from electroweak symmetry breaking \cite{Cohen:2020xca}, where a Higgs EFT (HEFT) approach \cite{Feruglio:1992wf} is more appropriate.  Recently, the most general allowed EFT operators in HEFT have been determined by considering the independent set of on-shell scattering amplitudes \cite{Shadmi:2018xan, Durieux:2019eor, Durieux:2020gip,Dong:2022jru,Liu:2023jbq, Chang:2022crb, Bradshaw:2023wco,Arzate:2023swz}.  By using these operators, it is possible to look for CLFV model-independently in the following fashion.  

In terms of CLFV decays, starting with two body final states, only $\ell \to \ell' \gamma$ is allowed by $SU(3)_c\times U(1)_{em}$ invariance.  On-shell, the flavor violating three point interactions of $\bar{\ell}' \ell \gamma$ are just the electric and magnetic dipole moment transitions (e.g.~\cite{Okada:1999zk,Durieux:2019eor,Chang:2022crb}).  This simplicity, is due to the fact that all of the other three point interactions can be related to higher point interactions through the use of equations of motion or integration by parts.  However, three body CLFV decays have a much more interesting structure, where $\ell \to \ell' \gamma \gamma,\; \ell' g g, \; \ell' \nu \nu, \; \bar{\ell}' \ell'' \ell'''$ are allowed by the symmetries.\footnote{For even higher body decays, there are also some recent studies, e.g.~\cite{Knapen:2023iwg}.}  The decays involving gluons are complicated by the closeness of the tau and muon masses to the QCD scale, but can lead to interesting decays into hadrons, such as $\tau\to e \pi^0$, which have constraints at $\sim 10^{-7}$ level \cite{ParticleDataGroup:2024cfk}.   For two photons, the branching ratio for $\mu \to e \gamma\gamma$ is currently constrained at $7.2\cdot 10^{-11}$ \cite{Fortuna:2022sxt, Bolton:1988af}.  There are no planned experimental improvements on such decays, although there are indirect constraints from $\mu\to e$ conversion and $\ell\to \ell' \gamma$ \cite{Davidson:2020ord, Fortuna:2023paj}.  The decays with neutrinos occur in the Standard Model and could be constrained by  Michel parameter extractions (e.g.~\cite{Pich:2013lsa,Rouge:2000um, Marquez:2022bpg}), which we'll return to later. 

In this paper, we will focus on decays of the muon and the tau, to three charged leptons:  
\begin{gather*} 
    \tau \rightarrow \bar{\mu}\mu\mu, \text{ }
    \tau \rightarrow  \bar{e}\mu\mu, \text{ }
    \tau \rightarrow \bar{\mu}ee,\text{ }
    \tau \rightarrow \bar{e}ee,\text{ }
    \mu  \rightarrow \bar{e}ee,\text{ }
    \tau \rightarrow \bar{\mu}\mu e, \text{ }
    \tau \rightarrow \bar{e}e \mu.
\end{gather*}
Past experiments have set an upper bound on the three body decays of the tau at $\text{BR}(\tau \rightarrow \bar{f}f'f'') \leq 10^{-8}$ and for the muon at $\text{BR}(\mu \rightarrow \bar{e}ee) \leq 10^{-12}$, while upcoming searches, particularly Mu3e and Belle II, expect to push these branching ratios to $10^{-10}$ and $10^{-15}$, respectively \cite{SINDRUM:1987nra, Hayasaka:2010np, Belle-II:2022cgf, Mu3e:2020gyw}. As we demonstrate below, these improvements will enable sensitivity to higher dimension operators in the effective theory. Should these decays be observed, the corresponding operators may be deduced from the kinematic distribution of the daughter particles. This would enable us to determine which SM extensions are consistent with the observed low energy behavior. As we will show, constraints from unitarity and LEP2 suggest that operators up to dimension 7 are the most likely to be observed, so we will derive the differential decay rates as a function of the kinematic observables for the most general operators at mass dimension 6 and 7 in the effective theory. The kinematics of decays mediated through dimension 6 operators was previously explored in \cite{Okada:1999zk, PhysRevD.58.051901, Kuno:1999jp}. For comparison with the dimension 7 operators, we adopt different parametrizations for the polarized and unpolarized decays to emphasize the observable differences in decays mediated through operators of the different dimensions. 

The outline of this work is as follows:
%
In section 2, we demonstrate the ability of upcoming experimental searches, such as Belle II, Mu3e, and a future muon collider, to probe the kinematic signatures of operators at dimension 6 and higher, and comment on the constraints resulting from the LEP experiments and unitarity.  These results show that one should focus on operators of dimension 6 and 7.  
In sections 3 and 4, we calculate the scattering amplitude for spin averaged decays to indistinguishable and distinguishable final states. The resulting differential decay rates are plotted in section 5.
In section 6, we analyze polarized decays to distinguishable and indistinguishable final states, which have distinct angular distributions. The general decay amplitudes for these channels are calculated and the corresponding marginalized decay rates are plotted, showing that many of the degenerate kinematic predictions found in section 5 are broken by the angular information. We then include a simple statistics analysis of the various distributions to  estimate the number of observed decays needed to distinguish the underlying operator. For most operators and decay channels, this threshold is within the reach of upcoming searches.  In section 7, we adapt these results to $\ell \to \ell' \nu \bar{\nu}$ for the case of Dirac neutrinos, showing again that angular distributions are powerful.   
In section 8, we conclude and give future outlook.  
In Appendices A and B, we give formulas for the squared amplitudes for each  decay channel, allowing for nonzero final state masses.  
In Appendix C, we list all of the EFT operators for CLFV  decays to three leptons to all mass dimensions.

\section{Experimental Sensitivity to EFT Operators}
\label{sec:exptsensitivity}
In this section, we estimate the size of Wilson coefficients needed to get CLFV branching ratios that can be seen in next generation experiments.  We will also discuss the impact of unitarity, LEP, and a future muon collider in constraining the size of these coefficients.  We will break up the analysis for $\tau$ and $\mu$ decays.

As shown in Appendix C, the general CLFV decay amplitudes are mediated by operators of dimension 6 and higher.  In particular, in the nomenclature of \cite{Chang:2022crb}, there are both primary and descendant operators, where descendants are primary operators dressed by derivatives. These give descendant amplitudes which are primary amplitudes multiplied by Mandelstam invariants.  For $\ell \to \bar{\ell}' \ell'' \ell'''$ where $\ell''$ and $\ell'''$ are different flavors, there are only primary operators of dimension 6 and 7.  On the other hand, if $\ell''$ and $\ell'''$ are the same flavor and hence  indistinguishable, there are primary operators at 6, 7, 8, and 9.  

\subsection{$\tau$ Decays}
We let the decay of a tau to three charged leptons (and no other particles) be mediated through an operator of dimension $d$: $\frac{c}{v^{d - 4}}\mathcal{O}^{(d)} \subset \mathcal{L}_{int} $ with $v = 246  \text{ GeV}$ and $c$ a dimensionless constant. 

Estimating the decay rate:
\begin{align}
\Gamma_{\tau \rightarrow l_1l_2l_3} \approx \frac{1}{256 \pi^3}\frac{|c|^2}{v^{2d  - 8}}m_\tau^{2d - 7}. 
\end{align}
The corresponding branching ratio is then: 
\begin{align}
\text{BR}(\tau \rightarrow l_1l_2l_3) & \approx \frac{\Gamma_{\tau \rightarrow l_1l_2l_3}}{\Gamma_{tot}} \approx \frac{1}{2.2 \cdot 10^{-12} \text{ GeV}}\frac{1}{256 \pi^3}\frac{|c|^2}{v^{2d  - 8}}m_\tau^{2d - 7}\approx \frac {1.3\cdot 10^{25}} {140^{2 d}} | c | ^2,\\
c & \approx  2.6\cdot 10^{-18} (140)^d \sqrt{\frac{\rm{BR}(\tau\to \ell_1\ell_2\ell_3)}{10^{-10}}}.
\end{align}
Current constraints on the branching ratio for all combinations of daughter particles are set at an upper bound of $10^{-8}$ (90 \% CL) by Belle and BaBar, while future experiments (namely, Belle II) will push these bounds to $10^{-10}$ \cite{Belle-II:2022cgf}. Using $BR = 10^{-10}$, we estimate that the maximal value of $c$ accessible to experiments is:
\begin{align}
\text{BR}(\tau\to l_1 l_2 l_3) > 10^{-10}\quad \rightarrow \quad |c| \gtrsim 2\cdot 10^{-5},\; 3\cdot 10^{-3},\; 0.4,\; 50 \text{ for } d=6, 7, 8, 9.
\end{align}

Given our normalization choice where the Wilson coefficient is suppressed by factors of $v$, unitarity bounds roughly require new physical states with mass below $m \lesssim v/|c|^{1/(d-4)}.$  Thus, we expect that values of $|c| \gtrsim 1$ require new particles at the weak scale.  Given the lack of signals for particles at the electroweak scale at LEP and the LHC, this suggests that dimension 8 operators are borderline in terms of giving observable LFV tau decays at Belle II, while dimension 9 are unlikely to be seen.   

LEP data can also constrain some operators by considering $e^+e^- \rightarrow \mu^+\tau^-$ and  $e^+ e^- \rightarrow e^+ \tau^-$ production. 
Estimating
\begin{align}
\sigma(e\bar{e} \rightarrow \tau l) \approx \frac{1}{4\pi}\frac{|c|^2 E_{cm}^{2d - 10}}{v^{2d - 8}} \approx  0.858 (64 \pi^2) \frac{E_{cm}^{2d - 10} \text{BR}(\tau \rightarrow e\bar{e}l)\text{GeV}^3}{m_\tau^{2d - 7}}\; \text{fb}.
\end{align}
Using the integrated luminosity of a single LEP2 experiment of $\sim 0.6 $ fb$^{-1}$, and $E_{cm} \sim 200 \text{ GeV}$, we estimate the number of expected events at each dimension:
\begin{align}
\text{BR}(\tau\to \bar{e}e l) > 10^{-10}\quad \rightarrow \quad N_{LEP2}(\bar{e}e \to \tau \bar{l}) \gtrsim 8\cdot 10^{-5},\; 1,\; 10^4 \text{ for } d=6, 7, 8.
\end{align}
%
This firmly excludes operators of dimension 8 and higher, that  mediate  $\bar{e}e \rightarrow \tau l  $ interactions, from being of phenomenological interest in upcoming experiments. 

A future muon collider would be able to probe interactions involving muon anti-muon pairs: $\tau \rightarrow \bar{\mu}\mu e$ and $\tau\rightarrow \bar{\mu} \mu \mu$. We repeat the above calculation with the target center of mass energy of $10 \text{ TeV}$ and integrated luminosity of $ 10 \text{ ab}^{-1}$ (e.g.~\cite{AlAli:2021let,Black:2022cth,InternationalMuonCollider:2024jyv}) to evaluate the extent to which this future collider could probe such operators, giving:
\begin{align}
\text{BR}(\tau\to \bar{\mu}\mu l) > 10^{-10}\quad \rightarrow \quad N_{MuC}(\bar{\mu}\mu \to \tau \bar{l}) \gtrsim 3\cdot 10^3,\; 10^{11},\; 3\cdot 10^{18} \text{ for } d=6, 7, 8.
    \label{MuEst}
\end{align}
%
For the assumed branching ratio, the 10 TeV center of mass energy is larger than the mass scale suppressing the operators for $d=7,8$, making their corresponding predictions unreliable. However, this suggests that for those dimensionalities, the muon collider can directly produce the particles mediating the operators.  Given this caveat, the proposed muon collider would be capable of exploring any such operators accessible to upcoming decay searches. 

\subsection{$\mu \rightarrow \bar{e}ee$}
Repeating our above estimate for the decay of the muon: 
\begin{equation}
\Gamma_{\mu\to ee\bar{e}} \approx \frac{1}{256\pi^3} \frac{|c|^2 m_\mu^{2d_{\mathcal{O}_i}-7}}{v^{2d_{\mathcal{O}_i}-8}},
\end{equation}
giving

\begin{equation}
{\rm{BR}}(\mu\to ee\bar{e}) = \frac{\Gamma_{\mu\to ee\bar{e}}}{\frac{1}{192\pi^3} G_f^2 m_\mu^5} \approx \frac{3}{2}|c|^2 \left(\frac{m_\mu}{v}\right)^{2d_{\mathcal{O}_i}-12} \approx \frac{3.8\cdot 10^{40}}{2330^{2d}} |c|^2.
\end{equation}
This leads to 

\begin{equation}
c\approx 1.6\cdot 10^{-28} (2330)^d \sqrt{\frac{\rm{BR}(\mu\to ee\bar{e})}{10^{-15}}}.
\end{equation}
Evaluating the above at $\text{BR} = 10^{-15}$, the expected sensitivity of Mu3e, yields:  
\begin{align}
\text{BR}(\mu\to e e\bar{e})> 10^{-15}\quad \rightarrow \quad |c| \gtrsim 3\cdot 10^{-8},\; 6\cdot 10^{-5},\; 0.1,\; 300 \text{ for } d=6, 7, 8, 9.
\end{align}
%
This suggests that $d=6, 7$ are the least constrained by direct searches, since unitarity bounds on the higher dimension operators will likely lead to new states with mass below the electroweak scale. 

In addition, these operators can lead to LEP constraints from $\bar{e} e \to \bar{e} \mu$ production.  If we assume that the new states are above the LEP center of mass, we get a cross section 
\begin{align}
\sigma(\bar{e} e \to \bar{e} \mu) & \approx \frac{1}{4\pi}  \frac{|c|^2 {\rm{E}_{\rm{cm}}}^{2d_{\mathcal{O}_i}-10}}{v^{2d_{\mathcal{O}_i}-8}} \approx \frac{1}{6\pi}{\rm{BR}(\mu\to ee\bar{e})} \frac{m_\mu^{12-2d_{\mathcal{O}_i}} {\rm{E}_{\rm{cm}}}^{2d_{\mathcal{O}_i}-10}}{v^4},\\
& \approx 10^{-49} (1900)^{2d_{\mathcal{O}_i}} \left(\frac{\rm E_{cm}}{\rm 200 GeV}\right)^{2d_{\mathcal{O}_i}-10}  \frac{{\rm{BR}(\mu\to ee\bar{e})}}{10^{-15}} {\rm \; fb}.
\end{align}
which assuming $\rm{E}_{\rm{cm}}=200 \rm{\; GeV}$, $\rm{BR}(\mu\to ee\bar{e})=10^{-15}$, and $\sim 0.6$ fb$^{-1}$ of integrated luminosity recorded by each LEP2 experiment, we get
\begin{align}
\text{BR}(\mu\to \bar{e}e e) > 10^{-15}\quad \rightarrow \quad N_{LEP2}(\bar{e}e \to \mu \bar{e}) \gtrsim 10^{-10},\; 5\cdot 10^{-4},\; 2000 \text{ for } d=6, 7, 8.
\end{align}
which shows that $d= 6, 7$ are safe, but dimension 8 and higher are now firmly ruled out.   

Finally, if one makes the reasonable assumption that there is an operator with four electrons in addition to the one mediating $\mu \to 3e$ with the same coupling strength, then we can look for interference with the Standard Model $\bar{e}e\to \bar{e}e$.  There are bounds on dimension 6 operators for four electrons, with constraints on $\frac{g^2}{2\Lambda^2} \bar{e}_L \gamma_\alpha e_L \bar{e}_L \gamma^\alpha e_L$ of $\Lambda \gtrsim 10 \rm{\; TeV}$ for $g^2=4\pi$.  This places an approximate constraint

\begin{equation}
c \left(\frac{\rm{E}_{\rm{cm}}}{v}\right)^{d_{\mathcal{O}_i}-4}
\lesssim \frac{4\pi \rm{E}_{\rm{cm}}^2}{2 \Lambda^2}. 
\end{equation}
which for $d=6, 7, 8$ and $\Lambda = 10 \rm{\; TeV}$ gives $c\lesssim 4\cdot 10^{-3},\, 5\cdot 10^{-3},\, 6\cdot 10^{-3}$, which   is safe for $d=6$, is close to setting a constraint on the largest branching ratios for $d=7$, and rules out seeing any effect at Mu3e for $d=8$.

\section{Decays to Indistinguishable Final States}
We begin by examining the following decay processes: 
\begin{gather*} 
\tau \rightarrow \bar{\mu}\mu\mu, \text{ }
\tau \rightarrow  \bar{e}\mu\mu, \text{ }
\tau \rightarrow \bar{\mu}ee, \text{ }
\tau \rightarrow \bar{e}ee,\text{ }
\mu  \rightarrow \bar{e}ee,
\end{gather*}
where there are indistinguishable particles in the final state.  
At mass dimension 6, there are 6 independent operators which could mediate this decay, while there are 2 at dimension 7. We use the following basis in these calculations: 
\begin{align}
& \mathcal{O}_1^{(6)} = (\bar{f}P_R f')(\bar{f} P_R F), &   \nonumber \\
& \mathcal{O}_2^{(6)} = (\bar{f}P_L f')(\bar{f} P_L F), &  \nonumber \\
& \mathcal{O}_3^{(6)} = (\bar{f} \gamma^\alpha P_L f')(\bar{f}  \gamma_\alpha P_L F), \nonumber & \mathcal{O}_1^{(7)} = ([\partial_\alpha\bar{f}]P_L f')(\bar{f} \gamma^\alpha P_L F), \nonumber \\[-.35cm] & &  \\[-.35cm]
& \mathcal{O}_4^{(6)} = (\bar{f} \gamma^\alpha P_R f')(\bar{f}   \gamma_\alpha P_L F),  & \mathcal{O}_2^{(7)} = ([\partial_\alpha \bar{f}]P_R f')(\bar{f} \gamma^\alpha P_R F). \nonumber \\
& \mathcal{O}_5^{(6)} = (\bar{f}  \gamma^\alpha P_L f')(\bar{f}   \gamma_\alpha P_R F),  & \nonumber \\
& \mathcal{O}_6^{(6)} = (\bar{f}  \gamma^\alpha P_R f')(\bar{f}  \gamma_\alpha P_R F), \nonumber &  
\end{align}
The procedure for arriving at a complete basis of the operators is described in \cite{Chang:2022crb, Bradshaw:2023wco} and further discussion is in Appendix C, showing that there are also new operators at dimension 8 and 9. We parameterize the resulting amplitudes and differential decay rates in terms of the invariant masses of pairs of the final state particles:

\begin{equation}
m_{ij}^2 = (p_i + p_j)^2,\text{   } i, j \in \{1,2,3\},
\end{equation}
where we label the four-momenta of the decay products in the general decay as $F(p) \rightarrow \bar{f}'(p_1) f(p_2)f(p_3)$. The invariant masses are subject to the global constraint:

\begin{equation}
m_{12}^2 + m_{23}^2 + m_{13}^2 = \sum_{i = 1}^4 m_i^2.
\end{equation}
However, we will use all three in the amplitudes that follow, to emphasize the invariance under interchanging the indistinguishable particles $(m_{12}^2 \leftrightarrow m_{23}^2$) and to obtain more concise results. Approximating the final states as massless, the decay amplitudes for the above processes are identical. The masses of the daughter particles produce sub-percent level variations in the decay amplitudes. As such, we report the massless result below. Amplitudes  for each decay  calculated without neglecting the final state leptons masses are given in Appendix A.  \\

The spin- averaged decay amplitude resulting from a dimension 6 operator of the form (note that unlike the $c$'s, the $C_i$ coefficients have nonzero mass dimensions): 
\begin{align}
\mathcal{L}^{d = 6} = \sum^6_{i = 1} C_i \mathcal{O}^{(6)}_i +h.c. 
\end{align}
is
\begin{align}
|\overline{\mathcal{M}^{(d = 6)}}|^2  = & \frac{1}{2} \left( |C_1|^2 + |C_2|^2 + 16\left( |C_3|^2 + |C_6|^2 \right) \right) \left( \left( m_{12}^2 + m_{13}^2 \right) m_{23}^2 \right) \nonumber \\ 
& + \left( |C_4|^2 + |C_5|^2 \right) \left( 4 m_{12}^2 m_{13}^2 + 2 \left( m_{12}^2 + m_{13}^2 \right) m_{23}^2 \right).
\label{eq:dim6indist}
\end{align}
Similarly, for dimension 7, we have: 
\begin{equation}
|\overline{\mathcal{M}^{(d = 7)}}|^2  =  \frac{1}{2} \left( |C_1|^2 + |C_2|^2 \right) m_{23}^2 \left( m_{12}^4 + m_{23}^2 (m_{12}^2 + m_{13}^2) + m_{13}^4 \right).
\label{eq:dim7indist}
\end{equation}
Note that we don't provide the interference between dimension 6 and 7 amplitudes, since we assume that if the amplitudes appear at dimension 6, those contributions will dominate over dimension 7.

\section{Decays to Distinguishable Final States}

In the following decays, the daughter particles from the tau decay may be distinguished based on mass and/ or charge:
\begin{gather*}
\tau \rightarrow \bar{\mu}\mu e, \text{ }
\tau \rightarrow \bar{e}e\mu.
\end{gather*}
At dimension 6, there are now 10 operators which could mediate these decays, and 8 at dimension 7: 
\begin{align}
& \tilde{\mathcal{O}}_1^{(6)} = (\bar{l}_2 P_R \ell_1)(\bar{\ell}_3 P_R {\ell}), & \tilde{\mathcal{O}}_1^{(7)} =  \left(\bar{\ell}_2 P_R \partial_\alpha  {\ell}_1\right) \left( \bar{\ell}_3   \gamma^{\alpha } P_L {\ell}\right), 
 \nonumber  \\
& \tilde{\mathcal{O}}_2^{(6)} = (\bar{\ell}_2 P_L {\ell}_1)(\bar{\ell}_3 P_R {\ell}), & \tilde{\mathcal{O}}_2^{(7)} =  \left(\bar{\ell}_2 P_L \partial_\alpha  {\ell}_1\right) \left( \bar{\ell}_3   \gamma^{\alpha } P_L {\ell}\right),  \nonumber  \\
& \tilde{\mathcal{O}}_3^{(6)} = (\bar{\ell}_2 P_R {\ell}_1)(\bar{\ell}_3 P_L {\ell}), & \tilde{\mathcal{O}}_3^{(7)} =  \left(\bar{\ell}_2 P_R \partial_\alpha  {\ell}_1\right) \left( \bar{\ell}_3   \gamma^{\alpha } P_R {\ell}\right),  \nonumber   \\
& \tilde{\mathcal{O}}_4^{(6)} = (\bar{\ell}_2 P_L {\ell}_1)(\bar{\ell}_3 P_L {\ell}),  & \tilde{\mathcal{O}}_4^{(7)} =  \left(\bar{\ell}_2 P_L \partial_\alpha  {\ell}_1\right) \left( \bar{\ell}_3    \gamma^{\alpha } P_R {\ell}\right),  \nonumber   \\
& \tilde{\mathcal{O}}_5^{(6)} = (\bar{\ell}_2  \gamma_\alpha P_L {\ell}_1)(\bar{\ell}_3  \gamma^\alpha P_L {\ell}),  & \tilde{\mathcal{O}}_5^{(7)} =  \left(\bar{\ell}_2  \gamma^{\alpha }P_L {\ell}_1\right) \left( \bar{\ell}_3 P_R   \partial_\alpha {\ell}\right),   \nonumber   \\[-.35cm] \\[-.35cm]
& \tilde{\mathcal{O}}_6^{(6)} = (\bar{\ell}_2  \gamma_\alpha P_R {\ell}_1)(\bar{\ell}_3 \gamma^\alpha P_L{\ell}),  & \tilde{\mathcal{O}}_6^{(7)} =  \left(\bar{\ell}_2  \gamma^{\alpha }P_L {\ell}_1\right) \left( \bar{\ell}_3 P_L   \partial_\alpha {\ell}\right),   \nonumber  \\& \tilde{\mathcal{O}}_7^{(6)} = (\bar{\ell}_2  \gamma_\alpha P_L{\ell}_1)(\bar{\ell}_3  \gamma^\alpha P_R{\ell}),  & \tilde{\mathcal{O}}_7^{(7)} =  \left(\bar{\ell}_2  \gamma^{\alpha } P_R {\ell}_1\right) \left( \bar{\ell}_3 P_R   \partial_\alpha {\ell}\right),   \nonumber  \\
& \tilde{\mathcal{O}}_8^{(6)} = (\bar{\ell}_2  \gamma_\alpha P_R {\ell}_1)(\bar{\ell}_3  \gamma^\alpha P_R {\ell}),  & \tilde{\mathcal{O}}_8^{(7)} =  \left(\bar{\ell}_2   \gamma^{\alpha } P_R {\ell}_1\right) \left( \bar{\ell}_3 P_L   \partial_\alpha {\ell}\right).   \nonumber   \\
& \tilde{\mathcal{O}}_9^{(6)} = (\bar{\ell}_2 P_R \sigma_{\alpha \beta} {\ell}_1)(\bar{\ell}_3 P_R \sigma^{\alpha \beta} {\ell}),  &   \nonumber \\
& \tilde{\mathcal{O}}_{10}^{(6)} = (\bar{\ell}_2 P_L \sigma_{\alpha \beta} {\ell}_1)(\bar{\ell}_3 P_L \sigma^{\alpha \beta} {\ell}),   &
\nonumber 
\end{align}
Given the discussion in Section \ref{sec:exptsensitivity}, dimension 6 and 7 are the most likely to be of interest, given unitarity and collider constraints.  There are further operators at dimension 8 and higher, from dressing these operators with contracted derivatives.  
We note that although it appears swapping $\bar{\ell}_2 \leftrightarrow \bar{l}_3$ in the above would yield new operators, they are expressible as a linear combination of the above, due to the Fierz identities. Treating daughter particles as massless, the general spin- averaged amplitude for a decay of the form $\ell(p) \rightarrow \bar{\ell}_1(p_1) \ell_2(p_2) \ell_3(p_3)$ mediated exclusively through dimension 6 operators is:
\begin{align}
& |\overline{\mathcal{M}^{(d = 6)}}|^2 = \frac{1}{2} \left( |C_1|^2 + |C_2|^2 + |C_3|^2 + |C_4|^2 \right) \left( (M^2 - m_{12}^2) m_{12}^2 \right) 
+ 2 \left( |C_5|^2 + |C_8|^2 \right) (M^2 - m_{23}^2) m_{23}^2 \nonumber \\
& + 2 \left( |C_6|^2 + |C_7|^2 \right) m_{13}^2(m_{12}^2 + m_{23}^2) 
+ 8 \left( |C_{10}|^2 + |C_9|^2 \right) \left( -(m_{12}^2 + 2 m_{23}^2)^2 + M^2 (m_{12}^2 + 4 m_{23}^2) \right)\nonumber  \\
& + \text{Re} \left( C_1 C_9^* + C_4 C_{10}^* \right)(4 m_{12}^2 m_{13}^2)
\label{eq:dim6dist}
\end{align}
For decays through dimension 7 operators, we have: 
\begin{align}
|\overline{\mathcal{M}^{(d = 7)}}|^2 = & \frac{1}{2} \left( |C_1|^2 + |C_2|^2 + |C_3|^2 + |C_4|^2 \right) m_{13}^2 m_{12}^2(M^2 - m_{23}^2) \nonumber \\
&+ \frac{1}{2} \left( |C_5|^2 + |C_6|^2 + |C_7|^2 + |C_8|^2 \right) m_{13}^2 m_{23}^2(M^2 - m_{12}^2)
\label{eq:dim7dist}
\end{align}
Amplitudes for the above decays, computed without neglecting the masses of daughter particles, are shown in Appendix B.  Again, we do not list dimension 7 amplitudes interfering with dimension 6, since it is unlikely that dimension 7 dominates over nonzero dimension 6 effects.  
\begin{figure}[h]
 \includegraphics[width = 200pt]{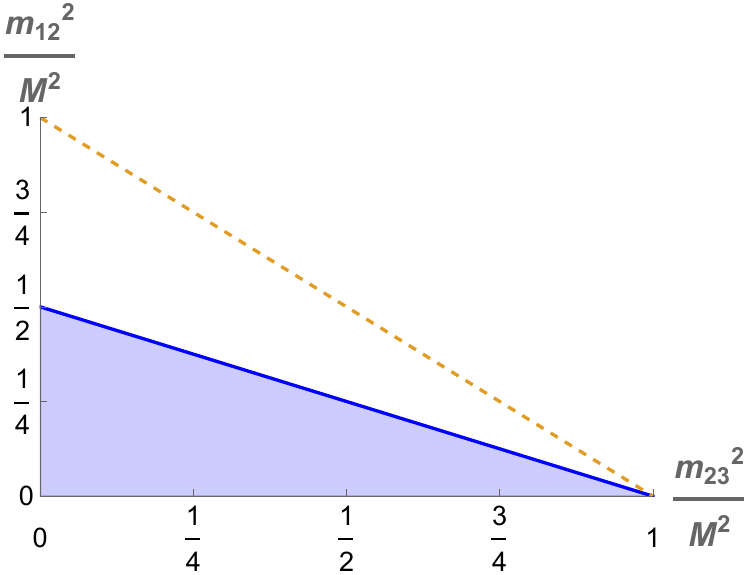}
 \centering
    \caption{\label{dalitz} Highlighting the domain of $(m_{23}^2, m_{12}^2)$ for the indistinguishable decays. If particles 2 and 3 were distinguishable, one would use the area under the dashed line.  Since they are indistinguishable, we distinguish them by requiring $m_{12}^2<m_{13}^2$, which results in the blue shaded region.  }
\end{figure}

\section{Unpolarized Differential Decay Rates}

For a $1 \rightarrow 3$ decay process, the differential decay rate may be expressed \cite{ParticleDataGroup:2024cfk}: 
\begin{equation} \label{spinsum}
\frac{d^2\Gamma_{F \rightarrow \bar{f}_1 f_2 f_3}}{dm_{12}^2 dm_{23}^2} = \frac{1}{256 \pi^3 M^3}|\overline{\mathcal{M}}|^2 
\end{equation}
To examine the dependence of the decay rate on either Lorentz invariant individually, we integrate over the domain of the other. 

{\bf  Indistinguishable case}:  If particles 2 and 3 are indistinguishable, there is an ambiguity between $m_{12}^2$ and $m_{13}^2$. For definiteness, we define particle 2 by requiring $m_{12}^2< m_{13}^2$. This requirement, along with the usual constraint of momentum conservation, yields the domain for $(m_{12}^2, m_{23}^2)$ depicted in Fig.~\ref{dalitz}.
To marginalize with respect to either Lorentz invariant, we then compute: 
\begin{equation}
\frac{d\Gamma_{F \rightarrow \bar{f}_1 f_2 f_3}}{dm_{23}^2}  =  \frac{1}{256 \pi^3 M^3} \int^{\frac{M^2 - m_{23}^2}{2}}_0  |\overline{\mathcal{M}}|^2 dm_{12}^2 
\end{equation}
\begin{equation}
\frac{d\Gamma_{F \rightarrow \bar{f}_1 f_2 f_3}}{dm_{12}^2}  =  \frac{1}{256 \pi^3 M^3}\int^{M^2 - 2 m_{12}^2}_0  |\overline{\mathcal{M}}|^2 dm_{23}^2 
\end{equation}

\begin{figure}[t]
     \includegraphics[width= 0.49\linewidth]{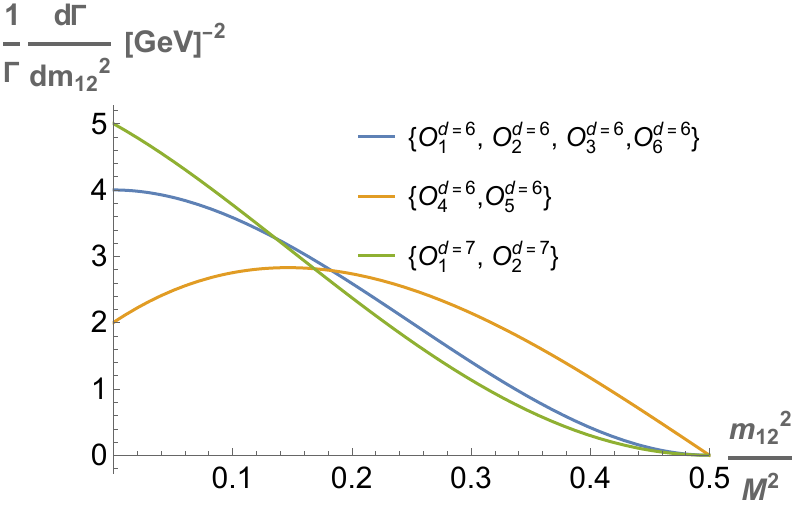}
     \;
     \includegraphics[width=0.49
 \linewidth]{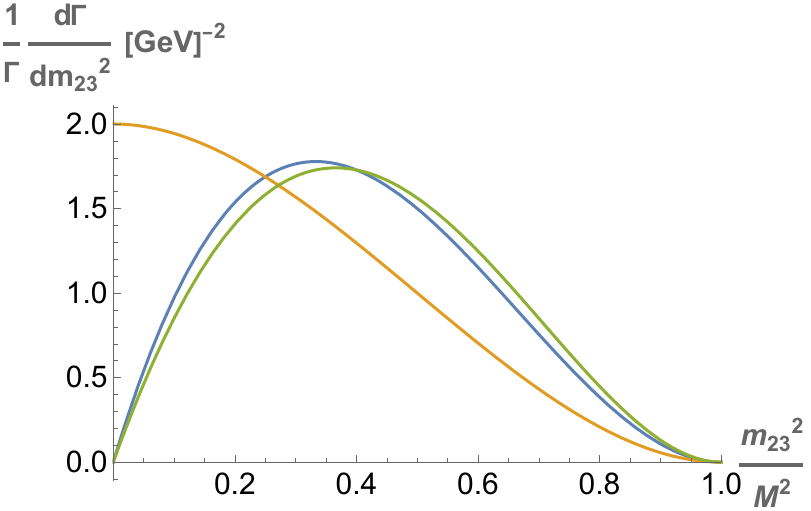}
     \centering
     \caption{Differential decay rates for $\tau$ and $\mu$ decays to 3 body final states with two indistinguishable particles: $\{\bar{e}ee\text{, } \bar{\mu}\mu\mu\text{, }\bar{\mu}ee\text{, } \bar{e}\mu\mu\}$. Masses of the daughter particles are neglected. The labeling of the invariant masses follows from enumerating the daughter particle's momenta as $F \rightarrow \bar{f}'(p_1)f(p_2)f(p_3)$, where we also require $m_{12}^2 \leq m_{13}^2$ to distinguish the two identical particles. $M$ is the rest mass of the decaying particle.}
\label{Indplots}
\end{figure}

\noindent
 Differential decay rates, resulting from individual operators in our basis mediating the decay, are unit normalized and displayed in Fig.~\ref{Indplots}. Variations in these plots between different final state particles are negligible, as the mass of the muon appears in powers of 2 or more in the single operator amplitudes. 

 As suggested by Eqns.~\ref{eq:dim6indist} and \ref{eq:dim7indist} and shown in the plots for $\frac{d\Gamma_{F \rightarrow \bar{f}_1 f_2 f_3}}{dm_{23}^2}$ and $\frac{d\Gamma_{F \rightarrow \bar{f}_1 f_2 f_3}}{dm_{12}^2}$, there are three families of operators which  are  distinguishable from one another (Fig.~\ref{Indplots}):
 \[\{\mathcal{O}^{d = 6}_1, \mathcal{O}^{d = 6}_2, \mathcal{O}^{d = 6}_3, \mathcal{O}^{d = 6}_6\}, \{ \mathcal{O}^{d = 6}_4, \mathcal{O}^{d = 6}_5\}, \text{ and } \{\mathcal{O}^{d = 7}_1, \mathcal{O}^{d = 7}_2\}.\]  For example, seeing the distribution of the first family only pins down the overall coefficient $|C_1^{(6)}|^2+ |C_2^{(6)}|^2+ 16 |C_3^{(6)}|^2+16 |C_6^{(6)}|^2$.
 Meanwhile, the polarized decay rates discussed below (and accessible to future searches), break these operators into six different families:
 \begin{equation*}
 \{\mathcal{O}^{d = 6}_1, \mathcal{O}^{d = 6}_6\}, \{\mathcal{O}^{d = 6}_2, \mathcal{O}^{d = 6}_3\}, \{ \mathcal{O}^{d = 6}_4 \}, \{ \mathcal{O}^{d = 6}_5 \}, \{\mathcal{O}^{d = 7}_1 \} \text{, and } \{\mathcal{O}^{d = 7}_2 \}
 \end{equation*}
Thus, polarized information breaks up most of the degeneracies, leaving just two pairs of operators that are indistinguishable.

\begin{figure}[t]
        \includegraphics[width=0.5 \linewidth]{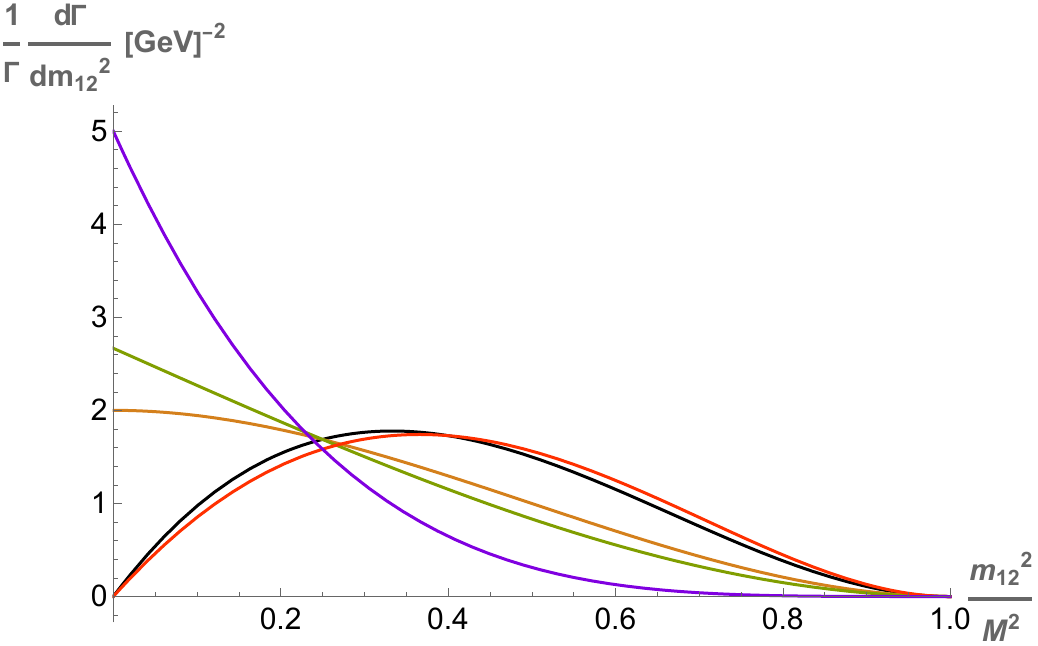} 
        \includegraphics[width=0.5 \linewidth]{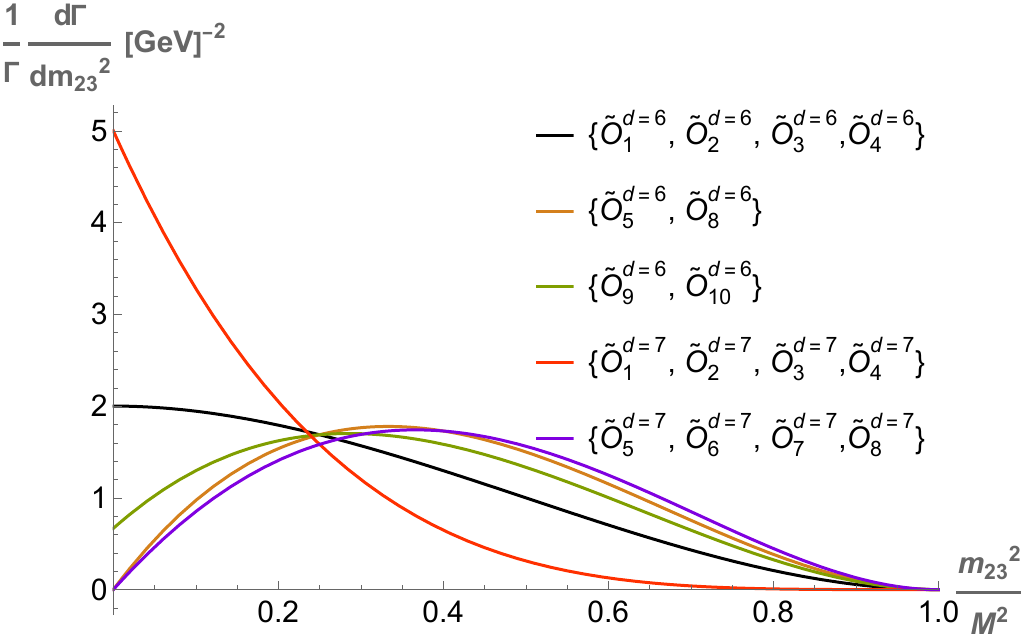}

\caption{Differential decay rates for a $\tau$ decaying to distinguishable final states: $\{\bar{\mu}\mu e\text{, }\bar{e}e\mu\}$. Masses of the daughter particles are neglected. The labeling of the invariant masses follows from enumerating the daughter particle's momenta as $\ell \rightarrow \bar{\ell_1}(p_1)\ell_2(p_2)\ell_3(p_3)$. $M$ is the rest mass of the decaying particle. $\mathcal{O}^{(d = 6)}_{\{6, 7\}}$ are not plotted, however the resulting differential decay rates are identical to those of  $\mathcal{O}^{(d = 6)}_{\{1, 2, 3,4\}}$ for $\frac{\text{d}\Gamma}{\text{d}m_{23}^2} $ and identical to those of $\mathcal{O}^{(d = 6)}_{\{5, 8\}}$ for $\frac{\text{d}\Gamma}{\text{d}m_{12}^2}$. \label{Distplots}
}
\end{figure}

{\bf  Distinguishable case}: For decays to distinguishable final states, the domain is constrained only by energy-momentum conservation: 
\begin{equation}
\frac{d\Gamma_{F \rightarrow \bar{f}_1 f_2 f_3}}{dm_{23}^2}  = \frac{1}{256 \pi^3 M^3} \int^{M^2 - m_{23}^2}_0  |\overline{\mathcal{M}}|^2 dm_{12}^2,
\end{equation}
\begin{equation}
\frac{d\Gamma_{F \rightarrow \bar{f}_1 f_2 f_3}}{dm_{12}^2}  = \frac{1}{256 \pi^3 M^3} \int^{M^2 - m_{12}^2}_0  |\overline{\mathcal{M}}|^2 dm_{23}^2 .
\end{equation}
 For distinguishable decays, the following families of operators yield identical distributions for $\frac{d\Gamma_{F \rightarrow \bar{f}_1 f_2 f_3}}{dm_{23}^2}$ (see Fig.~\ref{Distplots} and Eqns.~\ref{eq:dim6dist} and \ref{eq:dim7dist}.)
 \begin{equation*}
 \{\tilde{\mathcal{O}}^{d = 6}_1 , \tilde{\mathcal{O}}^{d = 6}_2 , \tilde{\mathcal{O}}^{d = 6}_3 , \tilde{\mathcal{O}}^{d = 6}_4, \tilde{\mathcal{O}}^{d = 6}_6, \tilde{\mathcal{O}}^{d = 6}_7 \}, \{\tilde{\mathcal{O}}^{d = 6}_5, \tilde{\mathcal{O}}^{d = 6}_8\},\{\tilde{\mathcal{O}}^{d = 6}_9, \tilde{\mathcal{O}}^{d = 6}_{10}\},
 \end{equation*}
 \begin{equation*}
 \{\tilde{\mathcal{O}}^{d = 7}_1 , \tilde{\mathcal{O}}^{d = 7}_2 , \tilde{\mathcal{O}}^{d = 7}_3 , \tilde{\mathcal{O}}^{d = 7}_4\},
 \{\tilde{\mathcal{O}}^{d = 7}_5 , \tilde{\mathcal{O}}^{d = 7}_6 , \tilde{\mathcal{O}}^{d = 7}_7 , \tilde{\mathcal{O}}^{d = 7}_8\}.
 \end{equation*}
 While the following families give the same distributions for  $\frac{d\Gamma_{F \rightarrow \bar{f}_1 f_2 f_3}}{dm_{12}^2}$ (Fig.~ \ref{Distplots}):
 \begin{equation*}
 \{\tilde{\mathcal{O}}^{d = 6}_1 , \tilde{\mathcal{O}}^{d = 6}_2 , \tilde{\mathcal{O}}^{d = 6}_3 , \tilde{\mathcal{O}}^{d = 6}_4\}, \{\tilde{\mathcal{O}}^{d = 6}_5,\tilde{\mathcal{O}}^{d = 6}_6, \tilde{\mathcal{O}}^{d = 6}_7, \tilde{\mathcal{O}}^{d = 6}_8\},\{\tilde{\mathcal{O}}^{d = 6}_9, \tilde{\mathcal{O}}^{d = 6}_{10}\},
 \end{equation*}
 \begin{equation*}
 \{\tilde{\mathcal{O}}^{d = 7}_1 , \tilde{\mathcal{O}}^{d = 7}_2 , \tilde{\mathcal{O}}^{d = 7}_3 , \tilde{\mathcal{O}}^{d = 7}_4\},
 \{\tilde{\mathcal{O}}^{d = 7}_5 , \tilde{\mathcal{O}}^{d = 7}_6 , \tilde{\mathcal{O}}^{d = 7}_7 , \tilde{\mathcal{O}}^{d = 7}_8\}.
 \end{equation*}
Combining the information in both distributions, makes $\{\tilde{\mathcal{O}}^{d = 6}_6, \tilde{\mathcal{O}}^{d = 6}_7\}$ a separate family.  

Polarized decay rates discussed below and potentially accessible to future searches, break these operators into eight different families:
 \begin{equation*}
 \{\tilde{\mathcal{O}}^{d = 6}_1, \tilde{\mathcal{O}}^{d = 6}_2, 
\tilde{\mathcal{O}}^{d = 6}_7 \},
\{\tilde{\mathcal{O}}^{d = 6}_3, \tilde{\mathcal{O}}^{d = 6}_4, 
\tilde{\mathcal{O}}^{d = 6}_6 \},
\{\tilde{\mathcal{O}}^{d = 6}_5, \tilde{\mathcal{O}}^{d = 7}_1, 
\tilde{\mathcal{O}}^{d = 7}_2 \},
\{\tilde{\mathcal{O}}^{d = 6}_8, \tilde{\mathcal{O}}^{d = 7}_3, 
\tilde{\mathcal{O}}^{d = 7}_4 \},
\{\tilde{\mathcal{O}}^{d = 6}_9 \},
\{\tilde{\mathcal{O}}^{d = 6}_{10} \},
\end{equation*}
\begin{equation*}
\{\tilde{\mathcal{O}}^{d = 7}_5, \tilde{\mathcal{O}}^{d = 7}_7 \}
 \text{, and } 
 \{\tilde{\mathcal{O}}^{d = 7}_6, \tilde{\mathcal{O}}^{d = 7}_8 \}.
\end{equation*}

\begin{SCfigure}
 \includegraphics[width=200 pt]{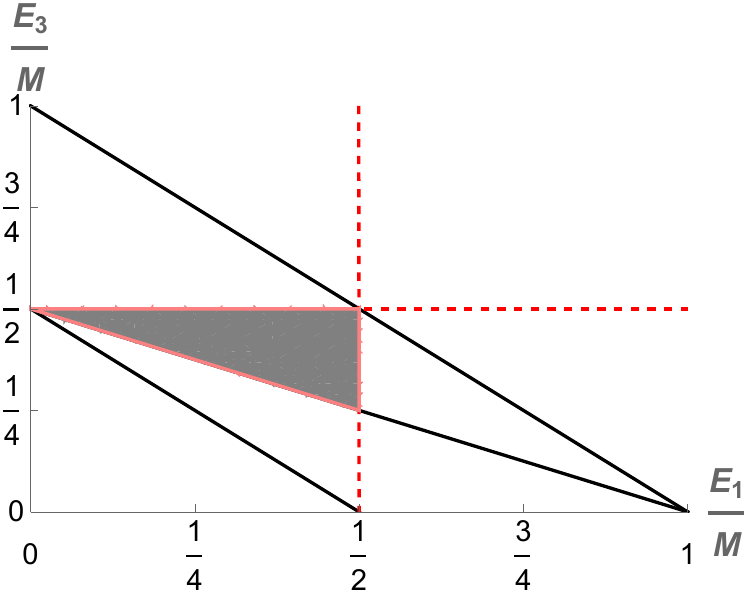}
  \caption{Plotting the constraints of the energies of the decay products for $\mu \rightarrow e^+e^-e^-$, where $E_1$ is the energy of the positron, $E_3$ is the energy of the electron with the highest energy. The shaded region corresponds to the domain of $(E_1, E_3)$, while $M$ is the mass of the muon.}
  \label{EnergyDomain}

\end{SCfigure}

\begin{figure}[h]

        \includegraphics[width= .48 \linewidth]{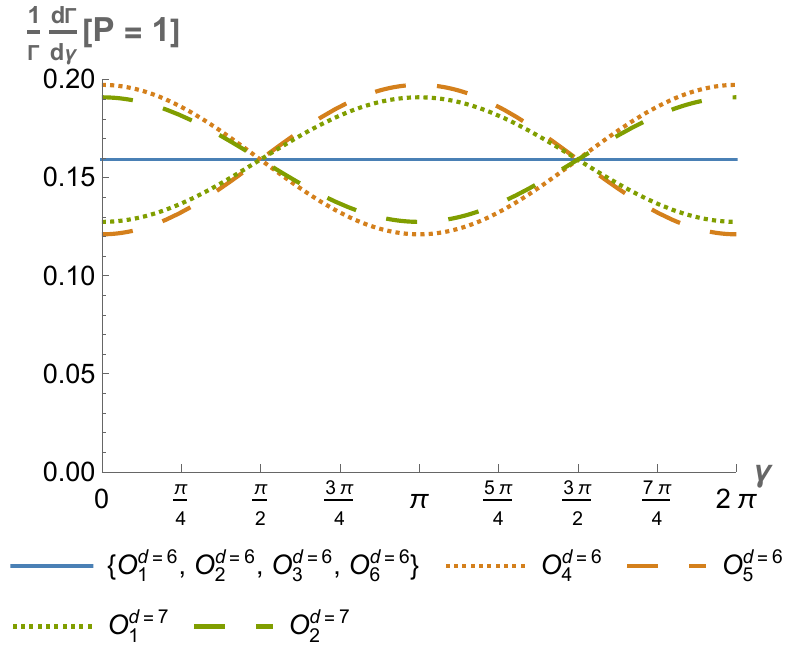}
        \quad
        \includegraphics[width= .48 \linewidth]{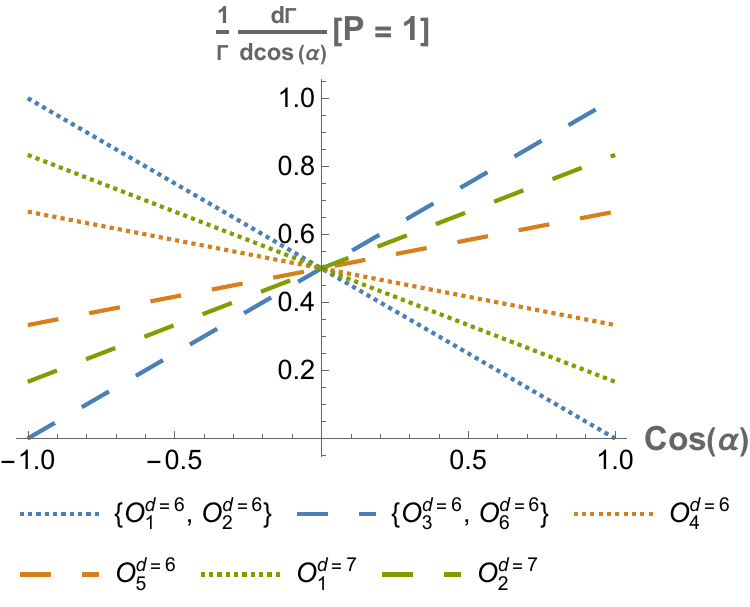}
        
        \label{fig:dGda, fig:dGdg}

\caption{Differential decay rates for 100\% polarized $\mu \rightarrow \bar{e}ee$ decays, as a function of the angular variables. As the polarization is reduced to zero, both distributions become flat. Note, operators which yield identical differential decay rates in spin summed calculations may be phenomenologically distinguished using the above. 
\label{fig:image3}
}
\end{figure}

\begin{figure}[h]

        \includegraphics[width= .54 \linewidth]{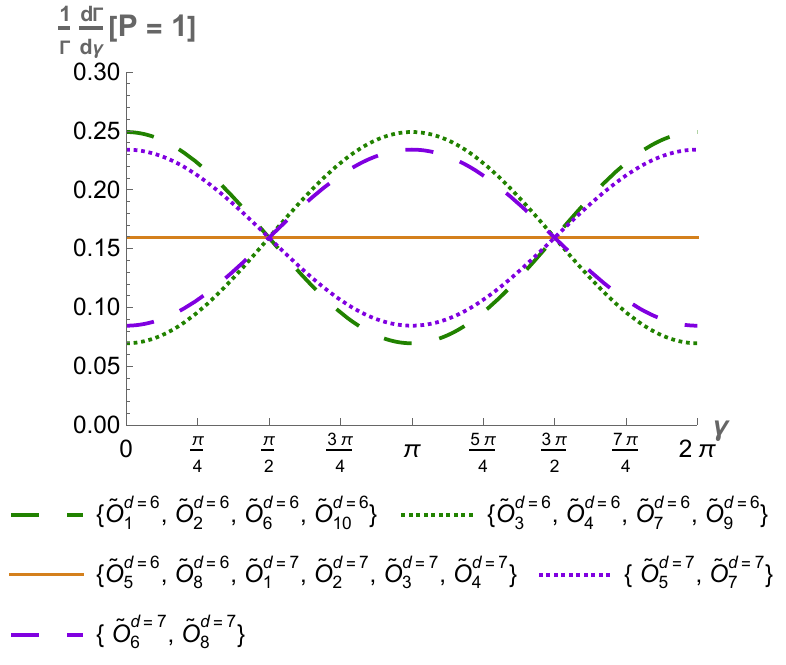}
        \quad
        \includegraphics[width= .44 \linewidth]{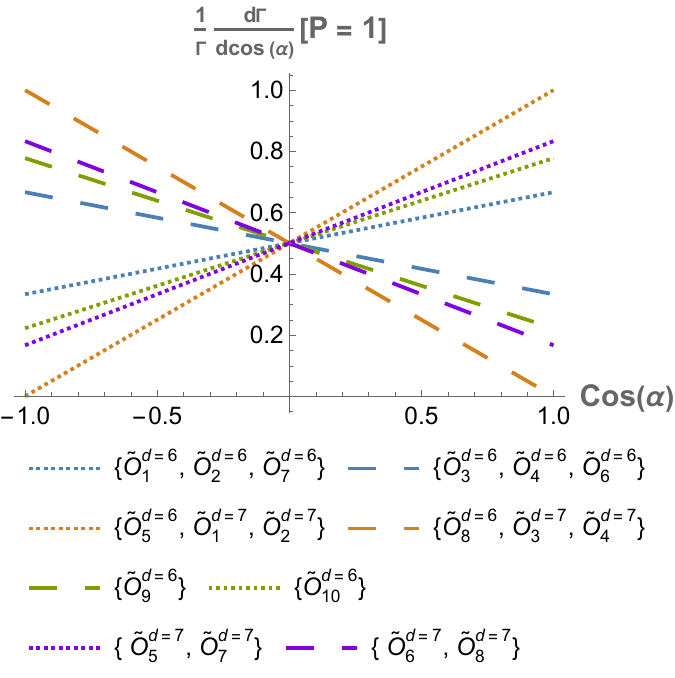}

\caption{Differential decay rates for 100\% polarized decays to distinguishable final states $(\tau \rightarrow \bar{\mu}\mu e \text{ and } \tau \rightarrow \bar{e} e \mu)$ decays, as a function of the angular variables. As the polarization is reduced to zero, both distributions become flat.  Note, operators which yield identical differential decay rates in spin summed calculations may be phenomenologically distinguished using the above. 
\label{fig:distPOL}
}
\end{figure}

\section{Polarized Differential Decay Rates}
The Mu3e experiment is expected to have a muon beam with  $\sim \! 90$\% polarization \cite{Maso:2023zjp}, while proposed upgrades to the Belle II experiment (Chiral Belle) would produce a $\sim \! 70\%$ polarization in the electron beam \cite{USBelleIIGroup:2022qro}, leading up to $70\%$ tau polarization . By assuming some spin orientation of the decaying particle, operators which yield the same differential decay rates in the spin-averaged calculation may be distinguished based on the angular dependence of their respective decay rates. 
By assuming some polarization, we lose two rotational symmetries and thus must introduce two new variables to parameterize the three 4-momenta of the daughter particles.

To describe the polarized decay kinematics, we will use the energy of the (distinguishable) positron $p_1^0 = |\vec{p_1}| = E_1$ and the energy of one of the (indistinguishable) electrons $p_3^0 = |\vec{p_3}| = E_3$. For massless final states, $m_{12}^2= M(M-2E_3)$ and $m_{13}^2 = M(M-2E_2)$, thus the earlier convention of $m_{12}^2 < m_{13}^2$ is equivalent to $E_3 > E_2.$
To parameterize the angular configuration, we let $(\alpha, \beta)$ be the angular coordinates specifying the orientation of  $p_1$ in the rest frame of the decaying muon, with the spin parallel (or antiparallel) to the $\hat{3}$ axis. Similarly, we let $(\theta, \gamma)$ denote the angular position of $p_3$ in a coordinate system in which $\frac{\vec{p_1}}{|\vec{p_1}|}$ defines the $\hat{3}'$ axis (again, in the rest frame of the decaying muon). We can then rotate back to the original (unprimed) frame via the rotation operator:
\begin{equation}
\mathcal{R}_{\hat{z}'}(\beta) \mathcal{R}_{\hat{y}'}(\alpha) =  \begin{pmatrix}
  1 & 0 & 0 &0\\ 
  0 & \cos{\beta} & -\sin{\beta} & 0 \\
  0 & \sin{\beta} & \cos{\beta} & 0 \\
  0 & 0 & 0 & 1\\
\end{pmatrix} 
\begin{pmatrix}
1 & 0 & 0 &0\\ 
  0 & \cos{\alpha} & 0 & \sin{\alpha}\\
  0 & 0 & 1 & 0 \\
  0 & -\sin{\alpha} & 0 & \cos{\alpha}\\
  \end{pmatrix}.
\end{equation}
Using our kinematic constraints, we may determine $\theta$ as a function of $E_1, E_3$:
\begin{equation}
\cos{\theta} = \frac{1}{E_1E_3} \left(\frac{M^2}{2} - M(E_1 + E_3) + E_1 E_3 \right).
\end{equation}
The momenta of our particles in the unprimed frame are then given:
\begin{equation}
p_1 = E_1
\begin{pmatrix}
1 \\
\sin{\alpha}\cos{\beta} \\ 
\sin{\alpha}\sin{\beta}\\
\cos{\alpha}\\
\end{pmatrix},  \; 
   p_3 = E_3
\begin{pmatrix}
1 \\
\sin{\theta}\cos{\alpha}\cos{\beta}\cos{\gamma} - \sin{\beta}\sin{\theta}\sin{\gamma} + \cos{\theta}\sin{\alpha}\cos{\beta} \\
\sin{\theta} \cos{\gamma}\cos{\alpha}\sin{\beta} + \cos{\beta}\sin{\theta}\sin{\gamma} + \cos{\theta}\sin{\alpha}\sin{\beta} \\
-\sin{\alpha}\sin{\theta}\cos{\gamma} + \cos{\alpha}\cos{\theta}
\end{pmatrix}, 
\end{equation}
\begin{equation}
p_2^\mu = p^\mu -p_1^\mu - p_3^\mu.
\end{equation} 
To compute the traces for the polarized decays, we may expand $u(p, S_z)\bar{u}(p, S_z)$ in the rest frame as:

\begin{equation}
u(p, \pm 1/2)\bar{u}(p, \pm 1/2) = \frac{M}{2}(\gamma^0 \pm \gamma^3 \gamma^5 + 1 \pm \sigma^{12})
\end{equation}
with  
$p^\mu = (M,0,0,0).$
For polarized decays, the differential decay rate may be expressed \cite{ParticleDataGroup:2024cfk}:
\begin{equation}
d \Gamma  = \frac{1}{512 \pi^5 M} |\mathcal{M}|^2 dE_1 dE_3 d\beta d(\cos \alpha) d\gamma
\end{equation}

We first consider polarized decays to indistinguishable final states with daughter particle masses treated as negligible. For these channels, the general differential decay rate mediated through a dimension 6 operator in the massless limit is given:
\begin{align}
& \frac{d^4\Gamma (P)}{dE_1dE_3d(\cos{\alpha})d\gamma} = \frac{1}{256 \pi^4 M}  \Big((| C_1 |^2 + 16 | C_6 |^2) E_1 M^2 (P \cos(\alpha) - 1) (2 E_1 - M)\nonumber \\
& - (| C_2 |^2 + 16 | C_3 |^2) E_1 M^2 (P\cos(\alpha) + 1) (2 E_1 - M) 
- 4 | C_4 |^2 M^2 
\big( 2 E_1^2 
-2 P E_3 \sin(\alpha) \cos(\gamma) \sin(\theta) (E_1 + 2 E_3 - M)\nonumber \\
& + P \cos(\alpha) \big(2  E_3 \cos(\theta) (E_1 + 2 E_3 - M) 
+ 2  E_1 (E_1 + E_3) -  E_1 M \big) \\
&+ E_1 (4 E_3 - 3 M) + (M - 2 E_3)^2 \big)
 - 4 | C_5 |^2 M^2 \big( 2 E_1^2 + 2 P E_3 \sin(\alpha) \cos(\gamma) \sin(\theta) (E_1 + 2 E_3 - M)\nonumber \\
& - P \cos(\alpha) \big( 2  E_3 \cos(\theta) (E_1 + 2 E_3 - M) + 2  E_1 (E_1 + E_3) -  E_1 M\big) 
+ E_1 (4 E_3 - 3 M) 
+ (M - 2 E_3)^2 \big) \big). \nonumber
\end{align}
For a decay through a general dimension 7 operator, we have: 
\begin{align}
& \frac{d\Gamma^4(P)}{dE_1dE_3d(\cos{\alpha})d\gamma} =  \frac{1}{256\pi^4 M}   \Big( |C_1|^2 M^3(M - 2 E_1) \big( - 2 P E_3 \sin(\alpha) \cos(\gamma) \sin(\theta) (E_1 + 2 E_3 - M) \nonumber \\ & + P \cos(\alpha) \left(2 E_3 \cos(\theta) (E_1 + 2 E_3 - M) + E_1 (2 E_3 - M) \right) 
- M (E_1 + 4 E_3) + 4 E_3 (E_1 + E_3) + M^2 \big)\\ & + |C_2|^2 M^3 (M - 2 E_1)\big( 2 P E_3 \sin(\alpha) \cos(\gamma) \sin(\theta) (E_1 + 2 E_3 - M) \nonumber \\ &  -P \cos(\alpha) \left(2 E_3 \cos(\theta) (E_1 + 2 E_3 - M) + E_1 (2 E_3 - M) \right)
- M (E_1 + 4 E_3) + 4 E_3 (E_1 + E_3) + M^2 \big) \Big).\nonumber
\end{align} 
\begingroup
\allowdisplaybreaks
As we should anticipate from the geometry of the decay, $|\mathcal{M}|^2 $ is independent of the azimuthal angle, $\beta$, hence a factor of $2\pi$ is acquired upon integrating over it.

We now consider the differential decay rates for channels with distinguishable final states, while again neglecting the masses of the daughter particles. Through a general dimension 6 operator, we find: 
\begin{align}
&\frac{d\Gamma^4(P)}{dE_1dE_3d(\cos{\alpha})d\gamma} = \frac{1}{256 \pi^4} M \Big( 
(|C_1|^2 +|C_2|^2) E_3 (2 E_3 - M) 
\big(-P\sin(\alpha) \cos(\gamma) \sin(\theta) +P\cos(\alpha) \cos(\theta) - 1 \big)\nonumber \\& 
- (|C_3|^2 + |C_4|^2) E_3 (2 E_3 - M) 
\big(-P \sin(\alpha) \cos(\gamma) \sin(\theta) +P\cos(\alpha) \cos(\theta) + 1 \big) \nonumber \\&
- 4 |C_5|^2 E_1 (2 E_1 - M) (P \cos(\alpha) + 1) 
- 4 |C_6|^2 (2 E_1 + 2 E_3 - M) 
\big(P \cos(\alpha) (E_1 + E_3 \cos(\theta)) + E_1 \nonumber \\& - E_3P\sin(\alpha) \cos(\gamma) \sin(\theta) + E_3 - M \big) \nonumber \\&
+ 4 |C_7|^2 (2 E_1 + 2 E_3 - M) 
\big(P \cos(\alpha) (E_1 + E_3 \cos(\theta)) - E_1 - E_3P\sin(\alpha) \cos(\gamma) \sin(\theta) - E_3 + M \big) \nonumber \\&
+ 4 |C_8|^2 E_1 (2 E_1 - M) (P \cos(\alpha) - 1) \nonumber \\&
+ 16 |C_9|^2 
\big(-8 E_1^2 - E_3P\sin(\alpha) \cos(\gamma) \sin(\theta) (4 E_1 + 2 E_3 - M) 
+P\cos(\alpha) \big(E_3 \cos(\theta) (4 E_1 + 2 E_3 - M) \nonumber \\& + 4 E_1 (2 E_1 + E_3 - M) \big) 
- 8 E_1 E_3 + 8 E_1 M - 2 E_3^2 + 5 E_3 M - 2 M^2 \big) \nonumber \\&
- 16 |C_{10}|^2 
\big(8 E_1^2 - E_3P\sin(\alpha) \cos(\gamma) \sin(\theta) (4 E_1 + 2 E_3 - M) 
+P\cos(\alpha) \big(E_3 \cos(\theta) (4 E_1 + 2 E_3 - M) \nonumber \\& + 4 E_1 (2 E_1 + E_3 - M) \big) 
+ 8 E_1 E_3 - 8 E_1 M + 2 E_3^2 - 5 E_3 M + 2 M^2 \big) \nonumber \\&
- 8 \operatorname{Re}(C_1 C_9^*) 
\big(E_3P\sin(\alpha) \cos(\gamma) \sin(\theta) (2 E_1 + 2 E_3 - M) 
- E_3P\cos(\alpha) \big(\cos(\theta) (2 E_1 + 2 E_3 - M) + 2 E_1 \big) 
\nonumber \\& + (2 E_3 - M) (2 E_1 + E_3 - M) \big) 
 \nonumber \\&
- 8 \operatorname{Re}(C_4 C_{10}^*) 
\big(-E_3P\sin(\alpha) \cos(\gamma) \sin(\theta) (2 E_1 + 2 E_3 - M) 
+ E_3P\cos(\alpha) \big(\cos(\theta) (2 E_1 + 2 E_3 - M) + 2 E_1 \big) \nonumber \\&
+ (2 E_3 - M) (2 E_1 + E_3 - M) \big) \nonumber \\&
+ 16\operatorname{Im}(C_1 C_9^* +C_4 C_{10}^* )E_1 E_3  P \sin(\alpha) \sin(\gamma) \sin(\theta)\Big)
\end{align}
\noindent
While for a general dimension 7 operator, the differential decay rate is:

\begin{align}
& \frac{d\Gamma^4(P)}{dE_1dE_3d(\cos{\alpha})d\gamma} =\frac{1}{256 \pi^4} M^2 \big(2 (E_1 + E_3) - M \big) 
\Big( 
 \big(|C_1|^2 + |C_2|^2 \big) E_1 (P \cos(\alpha) + 1) (M-2 E_3) \nonumber \\&
+ \big(|C_3|^2 + |C_4|^2 \big) E_1 (-P\cos(\alpha) + 1) (M-2 E_3 ) \nonumber \\&
+ \big(|C_5|^2 + |C_7|^2 \big) E_3 (M-2 E_1) 
\big(P\sin(\alpha) \cos(\gamma) \sin(\theta) - P \cos(\alpha) \cos(\theta) + 1 \big) \nonumber \\&
+ \big(|C_6|^2 + |C_8|^2 \big) E_3 (M-2 E_1) 
\big(-P \sin(\alpha) \cos(\gamma) \sin(\theta) + P \cos(\alpha) \cos(\theta) + 1 \big) 
\Big)
\end{align}
\endgroup

\subsection{Plotting Polarized Decay Rates}
To examine the differential dependence of the decay rate on the angular parameters, we may marginalize over the energy variables and one of the angular variables, by integrating over their respective domains (instead marginalizing over both angular variables recovers the spin-summed results above). The angular variables satisfy $\alpha \in [0, \pi]$ and  $\gamma \in [0, 2\pi]$. By momentum conservation (in the massless limit) we have $0 \leq E_1, E_3 \leq \frac{M}{2}$. Energy conservation yields the additional constraint: $\frac{M}{2} \leq E_1 + E_3 \leq M$. For decays to indistinguishable final states, we enforce the condition  $E_3 \geq E_2$ to remove ambiguity in our parametrization of the identical particles $2$ and $3$. This is equivalent the requirement $m_{12}^2 < m_{23}^2$, imposed earlier. Given that $E_1 + E_2 = M - E_3$, this implies $E_3 \geq \frac{M - E_1}{2}$. We summarize these constraints in Fig.~\ref{EnergyDomain}. Given these constraints, our marginalized differential decay rates for indistinguishable decay channels take the form:

\begin{equation}
\frac{d\Gamma}{d\gamma} =\int^\pi_0 \sin(\alpha) d\alpha \int^{\frac{M}{2}}_{\frac{M}{4}} dE_3 \int_{M - 2 E_3}^\frac{M}{2} dE_1 \frac{d^4\Gamma(P, E_1, E_3, \gamma, \alpha)}{dE_1dE_3d(\cos{\alpha})d\gamma},
\end{equation}
\begin{equation}
\frac{d\Gamma}{d\cos(\alpha)} =\int^{2\pi}_0  d\gamma \int^{\frac{M}{2}}_{\frac{M}{4}} dE_3 \int_{M - 2 E_3}^\frac{M}{2} dE_1 \frac{d^4\Gamma(P, E_1, E_3, \gamma, \alpha)}{dE_1dE_3d(\cos{\alpha})d\gamma},
\end{equation}
with $P$ being the polarization of the parent particle. The equivalent formulas for distinguishable decays are found by integrating $E_1$ over $(\frac{M}{2} - E_3, \frac{M}{2})$ and $E_3$ over $(0, \frac{M}{2})$. 
Plotting the differential decay rates for each operator emphasizes how the phenomenology of a decay may be used to identify the underlying operators, see Fig.~\ref{fig:image3}. Importantly, operators that yield identical differential decay rates in the spin-averaged calculation are distinguishable based upon the dependence of their resulting polarized decay rates on $\alpha, \gamma$.  In particular, as the figures show, $\alpha$ has much more  power than $\gamma$ to distinguish between the different operators. As discussed earlier,  incorporating this angular information results in six families of degenerate operators for indistinguishable channels, and eight families for distinguishable decays, compared to the three and five (respectively) obtained from the distributions of $m_{12}^2$ and $m_{23}^2.$
\subsection{Quantifying the Distinguishability of EFT operators}
As shown in the previous subsection, there is useful information in the angular distributions that can distinguish different operators.  In this subsection, we will perform a simple analysis, quantifying the number of events needed to distinguish different operators using the $\cos(\alpha)$ distributions.  
The unit normalized differential decay rates, marginalized with respect to the other variables, take the general form: \\
\begin{equation}
\frac{d\Gamma}{d\cos(\alpha)} \equiv f(\cos(\alpha)) = \frac{1}{2}(1 + q \cos(\alpha))
\end{equation}
\\
where the value of $q$ is determined by the operator which mediates the decay as well as the lepton polarization. We consider how $q$ can be estimated, based upon experimental observation of $N$ total decays, corresponding to $i=1,\ldots, N$ measured values of $\cos(\alpha_i)$.  Treating $\cos(\alpha)$ as a random variable with distribution given by $f(\cos(\alpha))$, we find its expectation value to be: 
\begin{equation}
\langle \cos(\alpha)\rangle = \frac{q}{3}.
\end{equation}
Thus enabling $q$ to be estimated via $\hat{q} = 3\; \langle\cos(\alpha)\rangle$.

The uncertainty of $\hat{q}$ as a function of $N$ and the lepton polarization enables us to predict the number of observed decays required to distinguish between the various operators.  Using the $\cos(\alpha)$ distribution, we get
\begin{equation}
\sigma_{\hat{q}} = 3 \sqrt{\frac{\text{Var}(\cos(\alpha))}{N}} =  3  \sqrt{\frac{\langle \cos^2(\alpha)\rangle - \langle \cos(\alpha)\rangle^2}{N}} = \frac{\sqrt{3 - q^2}}{\sqrt{N}}.
\end{equation}
If we consider only the observed values of $\cos(\alpha_i)$ (and no other information), we may predict the number of decays necessary to distinguish between the distinct families in Fig.~\ref{fig:image3}.  To estimate the minimum number of events $\hat{N}$ necessary to distinguish hypothesis $x$ versus $y$, assuming  $y$ is true, we require that $\Delta q= |q_x-q_y| = 1.64 \sigma_{\hat{q}_y}$, corresponding to a $90 \%$ confidence level.
\subsubsection{Analyzing Decays to Indistinguishable Final States}
For decays to indistinguishable final states, we have:
\begin{equation}
q \in \left\{-P, \frac{-2P}{3}, \frac{-P}{3}, \frac{P}{3}, \frac{2P}{3}, P\right\} \text{ for } \left\{ \mathcal{O}_{\{1, 2\}}^{d = 6}, \mathcal{O}_{1}^{d = 7}, \mathcal{O}_4^{d = 6}, 
\mathcal{O}_5^{d = 6}, \mathcal{O}_{2}^{d = 7}, \mathcal{O}_{\{3, 6\}}^{d = 6} \right\}
 \end{equation}
where $P$ is the beam polarization. The number of events needed to distinguish between the smallest slope difference, $\hat{N}$ is given by  $1.64\sigma_{\hat{q}} = \frac{P}{3}$. Substituting our results, we find: 
\begin{equation}
\hat{N} = \left(\frac{4.92}{P}\right)^2(3 - q^2)
\end{equation}
To be conservative, we can set $q = 0$, giving a larger $\sigma_{\hat{q}}$ which at 90\% C.L. gives:
\begin{equation}
\hat{N}(\Delta q = P/3) = \frac{72.6}{P^2}.
\label{eq:minN_ind}
\end{equation}
Note that this conservative bound also scales nicely with polarization.  
Thus for the projected beam polarization of $90\%$ at Mu3e experiment, we estimate a minimum of 90 decays to identify the family of operators mediating the decay, using only the observations of $\cos(\alpha_i)$ of the decaying particles.  On the other hand, using the $70\%$ maximum polarization at Chiral-Belle, would require 150 events to distinguish tau decay operators to indistinguishable final states. \\

Leveraging the energy distribution of the decay products enables us to reduce the number of events required to distinguish between these families. In light of Fig.~\ref{Indplots}, it is reasonable to suppose that we can distinguish $\{\mathcal{O}_4^{d = 6}, \mathcal{O}_5^{d = 6}\}$ from the two other degenerate families of operators $\{\mathcal{O}^{d = 6}_1, \mathcal{O}^{d = 6}_2, \mathcal{O}^{d = 6}_3, \mathcal{O}^{d = 6}_6\}$  and $ \{\mathcal{O}^{d = 7}_1, \mathcal{O}^{d = 7}_2\}$.  Separating between $\mathcal{O}_4^{d = 6}$ and $ \mathcal{O}_5^{d = 6}$, only requires enough events such that $1.64 \sigma_q < \frac{2}{3}P$, giving
\begin{equation}
\hat{N}(\Delta q = 2P/3) = \frac{18.15}{P^2}.
\end{equation}
On the other hand, the energy distributions of $\{\mathcal{O}^{d = 6}_1, \mathcal{O}^{d = 6}_2, \mathcal{O}^{d = 6}_3, \mathcal{O}^{d = 6}_6\}$  and $ \{\mathcal{O}^{d = 7}_1, \mathcal{O}^{d = 7}_2\}$ are similar and thus would be difficult to disentangle using energy.  However, the slope differences range from $P/3$ to $2P$, enabling some of the degeneracies to be broken in fewer decays than the general case Eq.~\ref{eq:minN_ind}.    

At Mu3e, given that the branching ratio sensitivity will improve by 3 orders of magnitude, we could expect at most about a thousand observed decays.  Thus, if the $90\%$ beam polarization can be achieved, this angular information could be used to distinguish various operators.  For tau decays, polarized beams for Belle-II  could produce polarized taus, leading to polarizations up to $70\%$.  With the tau branching ratio sensitivities improving by two orders of magnitude, we can expect up to a hundred observed decays, which given the above estimates, show that $\Delta q=P/3$ would be unlikely to be distinguished, but larger differences could be distinguished.  Of course, a multi-dimensional analysis using the Lorentz invariants, $\alpha$, and $\gamma$ variables will help optimize the distinguishing power beyond our one variable analysis.

\noindent
\subsubsection{Analyzing Decays to Distinguishable Final States}
For decays to distinguishable final states (Fig.~\ref{fig:distPOL}), we have:
\begin{equation*}
q \in \left\{-P, -\frac{2}{3}P, -\frac{5}{9}P, -\frac{1}{3}P, \frac{1}{3}P, \frac{5}{9}P, \frac{2}{3}P, P\right\},
\end{equation*}
\begin{equation*}
\text{ for } \left\{\{\tilde{\mathcal{O}}^{d = 6}_8, \tilde{\mathcal{O}}^{d = 7}_{\{3,4\}}\}, \tilde{\mathcal{O}}^{d = 7}_{\{6, 8\}}, \tilde{\mathcal{O}}^{d = 6}_9, \tilde{\mathcal{O}}^{d = 6}_{\{3, 4, 6\}},\tilde{\mathcal{O}}^{d = 6}_{\{1, 2, 7\}}, \tilde{\mathcal{O}}^{d = 6}_{10},  \tilde{\mathcal{O}}^{d = 7}_{\{5, 7\}}, \{\tilde{\mathcal{O}}^{d = 6}_5, \tilde{\mathcal{O}}^{d = 7}_{\{1,2\}}\}\right\}. 
\end{equation*}
Here the most challenging slope difference to distinguish is $P/9$.  Setting $1.64 \sigma_{\hat{q}} = \frac{P}{9}$, yields the condition:
\begin{equation}
\hat{N}(\Delta q = P/9) = \left(\frac{14.8}{P}\right)^2(3 - q^2)
\end{equation}
Setting $q = 0$ to minimize $q^2$, we find:
\begin{equation}
\hat{N}(\Delta q=P/9) = \frac{654}{P^2}.
\label{eq:minNdist}
\end{equation}
If we suppose a polarization of $\approx 70\%$ at Chiral-Belle, then we estimate $1330$ observations are needed to distinguish between any two families of operators.  Thus, Belle-II will not have enough events to distinguish this smallest slope difference.  
However, leveraging the unpolarized differential decay rates (Fig.~\ref{Distplots}) once again allows us to identify the underlying operator with fewer events.  From this kinematic information, we suppose that the mediating operator can be narrowed down to one of the following families: 
\begin{align*} &
 \{\tilde{\mathcal{O}}^{d = 6}_1 , \tilde{\mathcal{O}}^{d = 6}_2 , \tilde{\mathcal{O}}^{d = 6}_3 , \tilde{\mathcal{O}}^{d = 6}_4\}, \{ \{\tilde{\mathcal{O}}^{d = 6}_5, \tilde{\mathcal{O}}^{d = 6}_8\} , \{\tilde{\mathcal{O}}^{d = 6}_9, \tilde{\mathcal{O}}^{d = 6}_{10}\}\},\{\tilde{\mathcal{O}}^{d = 6}_6, \tilde{\mathcal{O}}^{d = 6}_7\}
\nonumber \\&
 \{\tilde{\mathcal{O}}^{d = 7}_1 , \tilde{\mathcal{O}}^{d = 7}_2 , \tilde{\mathcal{O}}^{d = 7}_3 , \tilde{\mathcal{O}}^{d = 7}_4\},
 \{\tilde{\mathcal{O}}^{d = 7}_5 , \tilde{\mathcal{O}}^{d = 7}_6 , \tilde{\mathcal{O}}^{d = 7}_7 , \tilde{\mathcal{O}}^{d = 7}_8\}.
\end{align*}
(While in principle $ \{\tilde{\mathcal{O}}^{d = 6}_5, \tilde{\mathcal{O}}^{d = 6}_8\}$ and $\{\tilde{\mathcal{O}}^{d = 6}_9, \tilde{\mathcal{O}}^{d = 6}_{10}\}$ are distinguishable, we group them together given their similar shapes for $\frac{\text{d}\Gamma}{\text{d} m_{12}^2}$ and $\frac{\text{d}\Gamma}{\text{d} m_{23}^2}$ which makes them difficult to separate in practice.)
Information from the angular distribution of the decay products enables the above families to be distinguished down to the following subsets: \noindent
\noindent
\begin{equation*}
 A: \{\tilde{\mathcal{O}}^{d = 6}_1 , \tilde{\mathcal{O}}^{d = 6}_2 , \tilde{\mathcal{O}}^{d = 6}_3 , \tilde{\mathcal{O}}^{d = 6}_4\} \rightarrow
  \begin{cases} 
       \{\tilde{\mathcal{O}}^{d = 6}_1 , \tilde{\mathcal{O}}^{d = 6}_2\} \\
       \{\tilde{\mathcal{O}}^{d = 6}_3 , \tilde{\mathcal{O}}^{d = 6}_4\}
   \end{cases}
 \text{       }
 B: \{ \{\tilde{\mathcal{O}}^{d = 6}_5, \tilde{\mathcal{O}}^{d = 6}_8\} , \{\tilde{\mathcal{O}}^{d = 6}_9, \tilde{\mathcal{O}}^{d = 6}_{10}\}\} \rightarrow
       \begin{cases} 
       \tilde{\mathcal{O}}^{d = 6}_5 \\
       \tilde{\mathcal{O}}^{d = 6}_8 \\
       \tilde{\mathcal{O}}^{d = 6}_9\\
       \tilde{\mathcal{O}}^{d = 6}_{10}
   \end{cases}
\end{equation*}
\begin{equation*}
 C: \{\tilde{\mathcal{O}}^{d = 6}_6, \tilde{\mathcal{O}}^{d = 6}_7\} \rightarrow
  \begin{cases} 
       \tilde{\mathcal{O}}^{d = 6}_6  \\
       \tilde{\mathcal{O}}^{d = 6}_7
   \end{cases}
   \text{        }
D: \{\tilde{\mathcal{O}}^{d = 7}_1 , \tilde{\mathcal{O}}^{d = 7}_2 , \tilde{\mathcal{O}}^{d = 7}_3 , \tilde{\mathcal{O}}^{d = 7}_4\} \rightarrow
  \begin{cases} 
       \{\tilde{\mathcal{O}}^{d = 7}_1 , \tilde{\mathcal{O}}^{d = 7}_2\} \\
       \{\tilde{\mathcal{O}}^{d = 7}_3 , \tilde{\mathcal{O}}^{d = 7}_4\}
   \end{cases}
\end{equation*}
\begin{equation*}
 E: \{\tilde{\mathcal{O}}^{d = 7}_5 , \tilde{\mathcal{O}}^{d = 7}_6 , \tilde{\mathcal{O}}^{d = 7}_7 , \tilde{\mathcal{O}}^{d = 7}_8\} \rightarrow
  \begin{cases} 
       \{\tilde{\mathcal{O}}^{d = 7}_5 , \tilde{\mathcal{O}}^{d = 7}_7\} \\
       \{\tilde{\mathcal{O}}^{d = 7}_6 , \tilde{\mathcal{O}}^{d = 7}_8\}
   \end{cases}
\end{equation*}
By substituting the appropriate minimum value of $\Delta q$ into our formula for $\hat{N}$, we estimate the number of events required to distinguish between the degenerate families in general, and assuming we have deduced the mediating operator to belong to one of the above subsets:

%
%
%
%
%
%
%
%
%
%
%
\begin{equation}
\hat{N}_A = \hat{N}_C= \frac{18.2}{P^2},\quad 
\hat{N}_B = \frac{40.8}{P^2}, \quad 
\hat{N}_D = \frac{2}{P^2}, \quad 
\hat{N}_E = \frac{4.5}{P^2}. 
\end{equation}
That is, even with poor polarization, a future experiment is capable of identifying the underlying decay operator up to two possibilities with the observation of less than 100 events, as shown for cases $A, C, D, E$. The novel dimension 7 operators explored in this work are especially good candidates to be identified, as a result of the distinguishability of their kinematic signatures both in their energy and angular distributions. In case $B$, the number of events needed to disseminate $\{ \tilde{\mathcal{O}}_{5}^{d = 6}, \tilde{\mathcal{O}}_{ 8}^{d = 6}, \tilde{\mathcal{O}}_{ 9}^{d = 6}, \tilde{\mathcal{O}}_{10}^{d = 6}\}$ down to a single operator is likely out of the reach of  upcoming searches, since it requires $\gtrsim 100$ events for $P\sim 0.7$. However, the number of events needed to reduce the set to two possible mediating operators $(\{ \tilde{\mathcal{O}}_{5}^{d = 6}, \tilde{\mathcal{O}}_{9}^{d = 6} \}$ or $\{\tilde{\mathcal{O}}_{ 8}^{d = 6}, \tilde{\mathcal{O}}_{10}^{d = 6}\})$  is lower by a factor of $6.25$. Again, a combined analysis of Lorentz invariants, $\alpha$, and $\gamma$ would have enhanced distinguishing power.  In fact, we've found that a similar one variable analysis using $\gamma$ alone can be more powerful than the $\alpha$ analysis, so this suggests substantial improvements could be obtained by a multi-dimensional analysis.

\section{Decays to neutrinos}
By modifying our results, we can also analyze polarized decays to a lepton and two neutrinos $\ell \rightarrow \bar{\nu} \ell' \nu$, assuming the neutrinos are Dirac so that $\bar{\nu}$ and $\nu$ are distinguishable. As the neutrinos are not experimentally observable, we parameterize our system such that $\alpha$ now corresponds to the angle between the outgoing lepton's trajectory and the initial lepton's polarization axis. Here we can use the operators for the charged lepton decay, by replacing $\ell_1 \to \nu, \bar{\ell}_2 \to \bar{\ell}', \bar{\ell}_3 \to \bar{\nu}.$  In the Standard Model, the $W$ generates $\mathcal{O}_5^{(d=6)}$ with $C_5^{d=6} = \frac{4 G_F}{\sqrt{2}}$ for the case where lepton flavor is preserved. \\

For a decay through a general dimension 6 EFT operator ($x=E_{\ell'}/(M/2)$):  \begin{align}
& \frac{d^2\Gamma (\ell\to \bar{\nu}\ell' \nu; P)}{dE_{\ell'}d(\cos{\alpha})} = \frac{M^4}{128 \pi^3}  \Big(\frac{1}{24}(| C_1 |^2 + | C_2 |^2+4|C_8|^2) x^2((3-2x) - P(1-2x) \cos(\alpha))\nonumber  \\ & 
+\frac{1}{24}(| C_3 |^2 + | C_4 |^2+4|C_5|^2) x^2((3-2x) + P (1-2x) \cos(\alpha))
\nonumber \\  & +|C_6|^2 x^2(1-x) (1 + P \cos \alpha) 
+ |C_7|^2 x^2(1-x) (1- P \cos \alpha)
  \nonumber \\ & 
+\frac{2}{3}|C_9|^2 x^2(15-14 x + P  (14x-13) \cos(\alpha))
+\frac{2}{3}|C_{10}|^2 x^2(15-14 x - P  (14x-13) \cos(\alpha))\nonumber \\ &  
+\frac{1}{3}\text{Re}(C_1 C_9^*)x^2(3-4x - P (5-4x) \cos \alpha)
+\frac{1}{3}\text{Re}(C_4 C_{10}^*)x^2(3-4x + P (5-4x) \cos \alpha) \Big). 
\end{align}
For a general dimension 7 operator: 
\begin{multline}
\frac{d^2\Gamma (\ell\to \bar{\nu}\ell' \nu; P)}{dE_{\ell'}d(\cos{\alpha})} = \frac{M^6}{128 \pi^3}  \Big(\frac{1}{24}(| C_1 |^2 + | C_2 |^2+| C_6 |^2 + | C_8 |^2) x^2 (1-x)(3-x -P (1+x) \cos (\alpha))\\+
\frac{1}{24}(| C_3 |^2 + | C_4 |^2+|C_5|^2+|C_7|^2) x^2 (1-x)(3-x + P (1+x) \cos (\alpha)) \Big).
\end{multline}

\begin{figure}[t]
 \includegraphics[width = .8 \linewidth]{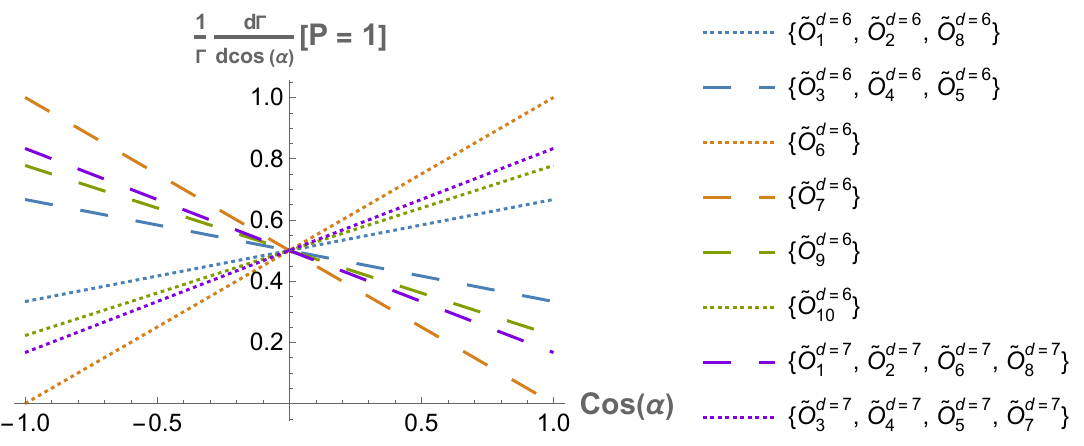}
 \centering
    \caption{\label{lvvcos} Differential decay rates as a function of $\cos(\alpha)$, where $\alpha$ is the angle of $\ell'$ about the polarization axis, in the decay $\ell \rightarrow \bar{\nu}\ell'\nu$. Due to the differing parametrizations of the daughter particles, the operators and their corresponding distributions are permuted relative to Fig.~\ref{fig:distPOL}.}
    \label{CosNuNu}
\end{figure}

\begin{figure}[t]
 \includegraphics[width = .9 \linewidth]{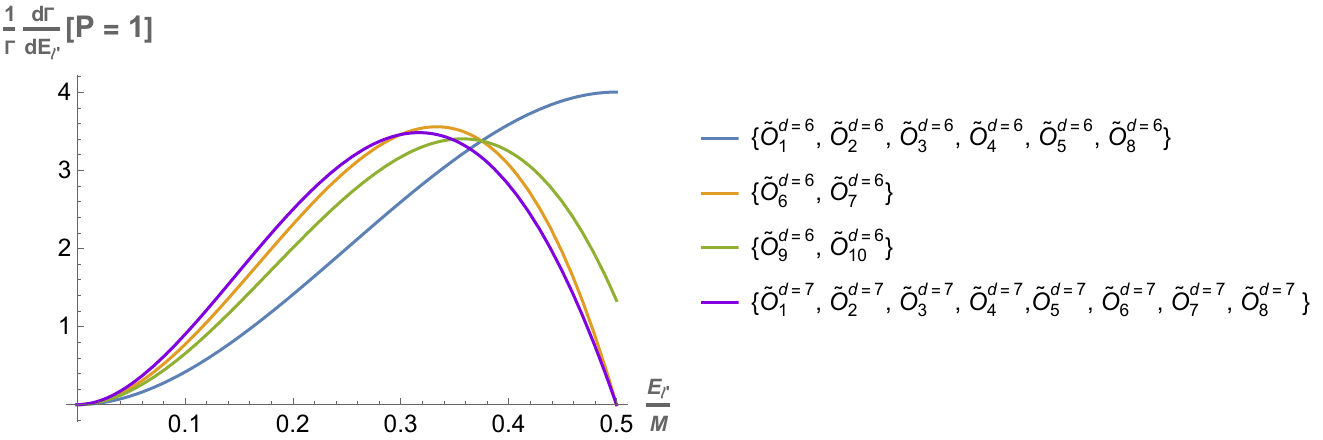}
 \centering
    \caption{\label{lvvEnergy} Differential decay rates as a function of the energy of the outgoing charged lepton, in the decay $\ell \rightarrow \bar{\nu}\ell'\nu$. }
\end{figure}

The distributions in $\cos(\alpha)$ and $E_{\ell'}$ are plotted in Figs.~\ref{lvvcos} and \ref{lvvEnergy}.  Again, one sees that the angular distributions are much more powerful than the energy ones to distinguish the different operators, leading to eight different families of operators:
\begin{align*}
& \{\tilde{\mathcal{O}}^{d = 6}_1 , \tilde{\mathcal{O}}^{d = 6}_2 , \tilde{\mathcal{O}}^{d = 6}_8\}, 
 \{\tilde{\mathcal{O}}^{d = 6}_3 , \tilde{\mathcal{O}}^{d = 6}_4 , \tilde{\mathcal{O}}^{d = 6}_5\},
  \{\tilde{\mathcal{O}}^{d = 6}_6\},
    \{\tilde{\mathcal{O}}^{d = 6}_7 \},\\ &
     \{\tilde{\mathcal{O}}^{d = 6}_9 \},
      \{\tilde{\mathcal{O}}^{d = 6}_{10} \},
       \{\tilde{\mathcal{O}}^{d = 7}_1 , \tilde{\mathcal{O}}^{d = 7}_2 , \tilde{\mathcal{O}}^{d = 7}_6,\tilde{\mathcal{O}}^{d = 7}_8\},       
       \{\tilde{\mathcal{O}}^{d = 7}_3 , \tilde{\mathcal{O}}^{d = 7}_4 , \tilde{\mathcal{O}}^{d = 7}_5,\tilde{\mathcal{O}}^{d = 7}_8\}.
 \end{align*}

\subsection{Quantifying Operator Distinguishability for Decays to Neutrinos}
In decays to  neutrinos, $\ell \rightarrow \bar{\nu}\ell'\nu$, we only observe the charged lepton in the final state. This limits the kinematic information at our disposal to  ${\frac{d \Gamma}{d\cos(\alpha)}}$ and $\frac{d \Gamma}{d\text{E}_{\ell'}}$ (see Fig.~\ref{CosNuNu}, \ref{lvvEnergy}).  From the former, we presume that the mediating operator may be narrowed into one of three families: 
\begin{equation*}
 \{\tilde{\mathcal{O}}^{d = 6}_1 , \tilde{\mathcal{O}}^{d = 6}_2 , \tilde{\mathcal{O}}^{d = 6}_3 , \tilde{\mathcal{O}}^{d = 6}_4, \tilde{\mathcal{O}}^{d = 6}_5, \tilde{\mathcal{O}}^{d = 6}_8\},
 \{\tilde{\mathcal{O}}^{d = 6}_9 , \tilde{\mathcal{O}}^{d = 6}_{10}\},
 \end{equation*}
 \begin{equation*}
 \{\{\tilde{\mathcal{O}}^{d = 6}_6, \tilde{\mathcal{O}}^{d = 6}_7 \}, \ \{\tilde{\mathcal{O}}^{d = 7}_1 , \tilde{\mathcal{O}}^{d = 7}_2, \tilde{\mathcal{O}}^{d = 7}_3 , \tilde{\mathcal{O}}^{d = 7}_4, \tilde{\mathcal{O}}^{d = 7}_5 , \tilde{\mathcal{O}}^{d = 7}_6,
       \tilde{\mathcal{O}}^{d = 7}_7 , \tilde{\mathcal{O}}^{d = 7}_8 \}\}
 \end{equation*}
While the latter family could in principle be separated into two families (delineated by the brackets), this is unlikely to be achieved in practice given the similarity of their corresponding  $\frac{d \Gamma}{d\text{E}_{\ell'}}$ functions. By observing the angular distribution of the $\ell'$ daughter particle, we may further distinguish the mediating operator from the above families down to the following subsets:
\begin{equation*}
 A: \{\tilde{\mathcal{O}}^{d = 6}_1 , \tilde{\mathcal{O}}^{d = 6}_2 , \tilde{\mathcal{O}}^{d = 6}_3 , \tilde{\mathcal{O}}^{d = 6}_4, \tilde{\mathcal{O}}^{d = 6}_5, \tilde{\mathcal{O}}^{d = 6}_8\} \rightarrow
    \begin{cases}
    \{\tilde{\mathcal{O}}^{d = 6}_1 , \tilde{\mathcal{O}}^{d = 6}_2 , 
     \tilde{\mathcal{O}}^{d = 6}_8\}\\
     \{\tilde{\mathcal{O}}^{d = 6}_3 , \tilde{\mathcal{O}}^{d = 6}_4 , 
     \tilde{\mathcal{O}}^{d = 6}_5\} 
    \end{cases}
            \text{            }
 B: \{\tilde{\mathcal{O}}^{d = 6}_9 , \tilde{\mathcal{O}}^{d = 6}_{10}\} \rightarrow
    \begin{cases}
    \tilde{\mathcal{O}}^{d = 6}_9 \\
    \tilde{\mathcal{O}}^{d = 6}_{10}
    \end{cases}
 \end{equation*}
 \begin{equation*}
 C: \{\{\tilde{\mathcal{O}}^{d = 6}_6, \tilde{\mathcal{O}}^{d = 6}_7 \}, \ \{\tilde{\mathcal{O}}^{d = 7}_1 , \tilde{\mathcal{O}}^{d = 7}_2, \tilde{\mathcal{O}}^{d = 7}_3 , \tilde{\mathcal{O}}^{d = 7}_4, \tilde{\mathcal{O}}^{d = 7}_5 , \tilde{\mathcal{O}}^{d = 7}_6,
       \tilde{\mathcal{O}}^{d = 7}_7 , \tilde{\mathcal{O}}^{d = 7}_8 \}\}
 \rightarrow
    \begin{cases}
    \{\tilde{\mathcal{O}}^{d = 7}_1 , \tilde{\mathcal{O}}^{d = 7}_2,    
    \tilde{\mathcal{O}}^{d = 7}_6 , \tilde{\mathcal{O}}^{d = 7}_8\}\\
    \{\tilde{\mathcal{O}}^{d = 7}_3 , \tilde{\mathcal{O}}^{d = 7}_4, 
    \tilde{\mathcal{O}}^{d = 7}_5 , \tilde{\mathcal{O}}^{d = 7}_7\} \\
    \tilde{\mathcal{O}}^{d = 6}_{6}\\
    \tilde{\mathcal{O}}^{d = 6}_7\\
    \end{cases}
 \end{equation*}
Compared to our earlier analysis from section 6.2, here there is a background from the Standard Model, which could be subtracted using a template.  However, to get an idea of what could be done, we {\bf assume} no background in the following.     The minimum number of events required to distinguish between the above families (and among the operators in general) up to a $90\%$ confidence level are reported below:
\noindent
\begin{equation}
\hat{N}_A =  \frac{18.2}{P^2},\quad 
\hat{N}_B = \frac{6.5}{P^2}, \quad 
\hat{N}_C = \frac{72.6}{P^2}. 
\end{equation}

\section{Conclusion}
An observation of charged lepton flavor violating interactions would be a clear indication of physics beyond the Standard Model, the nature of which can be identified or constrained based on its low energy behavior. To facilitate this, it is important to look for such interactions in a model-independent fashion, to enhance chances of discovery.  To do so, we have considered the most general  amplitudes for CLFV decays, showing that unitarity and constraints from LEP2 generally constrain dimension 8 and higher amplitudes from being seen at future experiments like Mu3e and Belle II.  

Thus, we have examined the decay kinematics which follow from all possible EFT operators at dimension 6 and 7, determining their  phenomenological signatures. In particular, operators at mass dimension 7 yield noticeably different differential decay rates compared to dimension 6 EFT operators. Operators which make degenerate predictions for unpolarized decays can be distinguished using the angular information of their polarized differential decay rates, with the exception of $\{\mathcal{O}^{d = 6}_1,\mathcal{O}^{d = 6}_6\}$ and $\{\mathcal{O}^{d = 6}_2,\mathcal{O}^{d = 6}_3\}$ . Upcoming experiments such as Mu3e and Belle II will have the capability to probe these dimension 7 operators (with Mu3e doing so with a polarized beam), where previous searches were limited by unitary bounds \cite{Mu3e:2020gyw, Belle-II:2022cgf}. To support this, we have performed a statistical analysis that estimates the number of events to break these degeneracies, which for many cases is accessible at future experiments.  Furthermore, the next generation of colliders will offer a window into the kinematics of interactions mediated by EFT operators of dimension 8 and beyond \cite{AlAli:2021let}, warranting eventual further exploration of the EFT operators which mediate CLFV interactions.  By modifying our results, we were also able to analyze the case of $\ell \to \ell' \bar{\nu} \nu$, showing that there is considerable information in the angular distribution of the emitted charged lepton, which could be constrained by Michel parameter fits (e.g.~\cite{Pich:2013lsa,Rouge:2000um, Marquez:2022bpg}).

In the future, it would be interesting to analyze other CLFV signals in the most general way, for instance $\mu$ to $e$ conversion (e.g.~\cite{Rule:2021oxe,Haxton:2022piv,Haxton:2024lyc}).  Also, it would be interesting to consider signals where there are new light states for CLFV decays, which has been discussed in the context of various models in \cite{Knapen:2023zgi, De:2024foq, Anselm:1985bp}.  By considering these general signals in a systematic fashion, the experimental prospects for discovering CLFV interactions can be maximized.        

\section*{Acknowledgements}
The work of S.C.~and T.D.~was supported in part by the U.S. Department of Energy under Grant Number DE-SC0011640. 

\section*{Appendix A: Decays to Indistinguishable Final States}
We consider the interaction:  $F(p) \rightarrow \bar{f}'(p_1)f(p_2)f(p_3)$, where $p_2^2 = p_3^2 = m^2$, $p_1^2 = m_1^2$, and $p^2 = M^2$. The general decay amplitude for this process, mediated through dimension 6 operators, is given: 
\begin{align}
&    |\overline{\mathcal{M}^{(d = 6)}}|^2 =  \frac{1}{2} \left(|C_1|^2 + |C_2|^2 + 16\left(|C_3|^2 + |C_6|^2\right)\right) \left((2m^2 - m_{12}^2 - m_{13}^2) (2m^2 - m_{23}^2)\right) \nonumber \\ &
+ \frac{1}{2} \left(|C_4|^2 + |C_5|^2\right) \big(4 \big(6m^4 + 2(m_1^2 - m_{12}^2)(m_1^2 - m_{13}^2) + (-2m_1^2 + m_{12}^2 + m_{13}^2)m_{23}^2\nonumber \\ & + m^2(4m_1^2 - 2(2(m_{12}^2 + m_{13}^2) + m_{23}^2))\big)\big) \nonumber \\ &
+ \text{Re}(C_1 C_3^* + C_2 C_6^*)(-8 M m_1 (-2m^2 + m_{23}^2)) 
+ \text{Re}(C_1 C_6^* + C_2 C_3^*)(8m^2(-2m^2 + m_{12}^2 + m_{13}^2))\nonumber \\ &
+ \text{Re}(C_1 C_4^* + C_2 C_5^*)(2m M (2m^2 + 2m_1^2 - m_{12}^2 - m_{13}^2))  \\ &
+ \text{Re}(C_1 C_5^* + C_2 C_4^*)(2m m_1 (-2m^2 - 2m_1^2 + m_{12}^2 + m_{13}^2 + 2m_{23}^2)) \nonumber \\ &
+ \text{Re}(C_1 C_2^*)(-4m^2 M m_1) 
+ \text{Re}(C_3 C_5^* + C_4 C_6^*)(8m M (2m^2 + 2m_1^2 - m_{12}^2 - m_{13}^2))\nonumber \\ &
+ \text{Re}(C_3 C_4^* + C_5 C_6^*)(8m m_1 (-2m^2 - 2m_1^2 + m_{12}^2 + m_{13}^2 + 2m_{23}^2)) 
+ \text{Re}(C_3 C_6^*)(-64m^2 M m_1)\nonumber \\ &
+ \text{Re}(C_4 C_5^*)(8M m_1 (-6m^2 + m_{23}^2)).\nonumber
\end{align}
For dimension 7 operators, we have:  
\begin{align}
&|\overline{\mathcal{M}^{(d = 7)}}|^2 = -\frac{1}{2} \left( |C_1|^2 + |C_2|^2 \right) 
\big( 4 m^6 + 2 m^4 \left( 8 m_1^2 - 3 (m_{12}^2 + m_{13}^2) - 4 m_{23}^2 \right) + m^2 \big( m_{23}^2 \left( 5 (m_{12}^2 + m_{13}^2) - 12 m_1^2 \right)  \nonumber \\ & + 2 \left( -2 m_1^2 + m_{12}^2 + m_{13}^2 \right)^2 + 2 m_{23}^4 \big) 
- m_{23}^2 \left( 2 m_1^4 - 2 m_1^2 (m_{12}^2 + m_{13}^2 + m_{23}^2) + m_{12}^4 
+ m_{23}^2 (m_{12}^2 + m_{13}^2) + m_{13}^4 \right) 
\big)  \nonumber \\ & + \text{Re}[C_1C_2^*] \left( 2 M m_1 (m_{23}^2 - 2 m^2)^2 - 12 m^4 M m_1 \right). 
\end{align}
As before, we don't present the interference between dimension 6 and 7 operators.  By replacing $m_1$ and $m$ with $m_e$ or $m_\mu$, one can get the finite mass results for 
\begin{gather*} 
\tau \rightarrow \bar{\mu}\mu\mu, \text{ }
\tau \rightarrow  \bar{e}\mu\mu, \text{ }
\tau \rightarrow \bar{\mu}ee, \text{ }
\tau \rightarrow \bar{e}ee,\text{ }
\mu  \rightarrow \bar{e}ee.
\end{gather*}

\section*{Appendix B: Decays to Distinguishable Final State Particles}
The general spin- averaged amplitude for a decay of the form $F(p^\mu) \rightarrow f_1(p_1^\mu) f_2(p_2^\mu) f_3(p_3^\mu)$ mediated exclusively through dimension 6 operators is:
\begingroup
\allowdisplaybreaks
\begin{align}
& |\overline{\mathcal{M}^{(d = 6)}}|^2 = \frac{1}{2} \left(|C_1|^2 + |C_2|^2 + |C_3|^2 + |C_4|^2\right) \left((m_1^2 - m_{12}^2 + m_2^2)(m_1^2 - m_{13}^2 + m_2^2 - m_{23}^2)\right) \nonumber \\ 
& + \frac{1}{2} \left(|C_5|^2 + |C_8|^2\right) \left(-4(m_{12}^2 + m_{13}^2 - m_2^2 - m_3^2)(m_2^2 - m_{23}^2 + m_3^2)\right) \nonumber \\
& + \frac{1}{2} \left(|C_6|^2 + |C_7|^2\right) \left(4(m_1^2 - m_{13}^2 + m_3^2)(m_1^2 - m_{12}^2 - m_{23}^2 + m_3^2)\right) \nonumber \\
& + \frac{1}{2} \left(|C_{10}|^2 + |C_9|^2\right) \big(16(m_1^4 - m_{13}^2 m_2^2 + m_2^4 + 4 m_{13}^2 m_{23}^2 - m_2^2 m_{23}^2 - 4(m_{13}^2 \nonumber  \\ & - m_2^2 + m_{23}^2)m_3^2 + 4 m_3^4 + m_{12}^2(m_{13}^2 - m_2^2 + m_{23}^2 - 4 m_3^2) - m_1^2(m_{12}^2 + m_{13}^2 + 2m_2^2 + m_{23}^2 - 4 m_3^2))\big) \nonumber \\ & 
+ \text{Re}(C_1 C_2^* + C_3 C_4^*) \left(2 m_1 m_2 (m_1^2 - m_{13}^2 + m_2^2 - m_{23}^2)\right)
+ \text{Re}(C_5 C_6^* + C_7 C_8^*) \left(4 m_1 m_2 (-m_1^2 + m_{13}^2 - m_2^2 + m_{23}^2)\right) \nonumber \\ & 
+ \text{Re}(C_9 C_{10}^*) \left(-192 M m_1 m_2 m_3\right)
+ \text{Re}(C_6 C_7^* + C_5 C_8^*) \left(-16 M m_1 m_2 m_3\right) \nonumber \\
&  + \text{Re}(C_2 C_3^* + C_1 C_4^*) \left(-4 M m_1 m_2 m_3\right)
+ \text{Re}(C_1 C_3^* + C_2 C_4^*) \left(-2 M (m_1^2 - m_{12}^2 + m_2^2) m_3\right)  \\
&  + \text{Re}(C_5 C_7^* + C_6 C_8^*) \left(4 M (m_1^2 - m_{12}^2 + m_2^2) m_3\right) \nonumber \\ &
+ \text{Re}(C_4 C_5^* + C_3 C_6^* + C_2 C_7^* + C_1 C_8^*) \left(2 m_2 m_3 (m_{12}^2 + m_{13}^2 - m_2^2 - m_3^2)\right) \nonumber \\
&  + \text{Re}(C_7 C_{10}^* + C_6 C_9^*) \left(-24 M m_2 (m_1^2 - m_{13}^2 + m_3^2)\right) \nonumber \\ &
+ \text{Re}(C_2 C_5^* + C_1 C_6^* + C_4 C_7^* + C_3 C_8^*) \left(-2 M m_2 (m_1^2 - m_{13}^2 + m_3^2)\right) \nonumber \\
&  + \text{Re}(C_5 C_{10}^* + C_8 C_9^*) \left(24 m_2 m_3 (-m_{12}^2 - m_{13}^2 + m_2^2 + m_3^2)\right) \nonumber \\
&  + \text{Re}(C_3 C_5^* + C_4 C_6^* + C_1 C_7^* + C_2 C_8^*) \left(2 m_1 m_3 (m_1^2 - m_{12}^2 - m_{23}^2 + m_3^2)\right) \nonumber \\
&  + \text{Re}(C_6 C_{10}^* + C_7 C_9^*) \left(24 m_1 m_3 (m_1^2 - m_{12}^2 - m_{23}^2 + m_3^2)\right) 
+ \text{Re}(C_8 C_{10}^* + C_5 C_9^*) \left(-24 M m_1 (m_2^2 - m_{23}^2 + m_3^2)\right) \nonumber \\
&  + \text{Re}(C_1 C_5^* + C_2 C_6^* + C_3 C_7^* + C_4 C_8^*) \left(2 M m_1 (m_2^2 - m_{23}^2 + m_3^2)\right) \nonumber \\
&  + \text{Re}(C_4 C_{10}^* + C_1 C_9^*) \big(4(m_1^4 + m_{12}^2(m_{13}^2 + m_2^2 - m_{23}^2) - m_1^2(m_{12}^2 + m_{13}^2 + m_{23}^2 - 2 m_3^2) \nonumber \\ & + m_2^2(m_{13}^2 - m_2^2 + m_{23}^2 - 2 m_3^2))\big).  \nonumber
\end{align}
\endgroup
For dimension 7 operators, we have: 
\begin{align}
&   |\overline{\mathcal{M}^{(d = 7)}}|^2 = \frac{1}{2} \left(|C_1|^2 + |C_2|^2 + |C_3|^2 + |C_4|^2\right)
\big(-\left(m_1^2 - m_{12}^2 + m_2^2\right) 
\big(m_1^4 + (m_{13}^2 - m_3^2)(m_{12}^2 + m_{13}^2  \nonumber \\ & - m_2^2 - m_3^2) 
- m_1^2 \left(m_{12}^2 +  2 m_{13}^2 - 2 m_2^2 + m_{23}^2 - m_3^2\right)\big)\big) \nonumber  \\
& + \frac{1}{2} \left(|C_5|^2 + |C_6|^2 + |C_7|^2 + |C_8|^2\right)
\big(\left(m_1^2 - m_{13}^2 + m_2^2 - m_{23}^2\right) 
\big(m_1^4 - m_{12}^2 m_2^2 - m_{13}^2 m_2^2 + m_2^4 - m_{13}^2 m_{23}^2 
- \nonumber \\ & m_1^2 \left(m_{12}^2 - m_2^2 + m_{23}^2\right) 
+ \left(m_{12}^2 + m_{13}^2 + m_{23}^2\right) m_3^2 - m_3^4\big)\big) \nonumber  \\
& - \operatorname{Re}\left(C_1 C_7^* + C_2 C_5^* + C_3 C_8^* + C_4 C_6^*\right)
\left(M m_2 \left(m_1^2 - m_{13}^2 + m_3^2\right)\left(-m_{12}^2 - m_{13}^2 + m_2^2 + m_3^2\right)\right) \nonumber  \\
& +\operatorname{Re}\left(C_1 C_5^* + C_2 C_7^* + C_3 C_6^* + C_4 C_8^*\right)
\left(M m_1 \left(m_1^2 - m_{13}^2 + m_3^2\right)\left(m_1^2 - m_{12}^2 - m_{23}^2 + m_3^2\right)\right) \nonumber  \\
& + \operatorname{Re}\left(C_1 C_6^* + C_2 C_8^* + C_3 C_5^* + C_4 C_7^*\right)
\left(m_1 m_3 \left(-m_{12}^2 - m_{13}^2 + m_2^2 + m_3^2\right)
\left(m_1^2 - m_{12}^2 - m_{23}^2 + m_3^2\right)\right) \nonumber  \\
& - \operatorname{Re}\left(C_1 C_8^* + C_2 C_6^* + C_3 C_7^* + C_4 C_5^*\right)
\left(m_2 m_3 \left(m_{12}^2 + m_{13}^2 - m_2^2 - m_3^2\right)^2\right) 
+ \operatorname{Re}\left(C_1 C_4^* + C_2 C_3^*\right)\left(-4 M m_1^3 m_2 m_3\right) \nonumber  \\
& + \operatorname{Re}\left(C_1 C_3^* + C_2 C_4^*\right)
\left(-2 M m_1^2 \left(m_1^2 - m_{12}^2 + m_2^2\right) m_3\right)   \\
& + \operatorname{Re}\left(C_1 C_2^* + C_3 C_4^*\right)
\Big(-2 m_1 m_2 \big(m_1^4 + (m_{13}^2 - m_3^2)\left(m_{12}^2 + m_{13}^2 - m_2^2 - m_3^2\right) 
\nonumber \\ &
- m_1^2 \left(m_{12}^2 + 2 m_{13}^2 - 2 m_2^2 + m_{23}^2 - m_3^2\right)\big)\Big) \nonumber  \\
& + \operatorname{Re}\left(C_5 C_8^* + C_6 C_7^*\right)
\left(4 M m_1 m_2 m_3 \left(m_1^2 - m_{12}^2 - m_{13}^2 + m_2^2 - m_{23}^2 + m_3^2\right)\right) \nonumber  \\
& + \operatorname{Re}\left(C_5 C_6^* + C_7 C_8^*\right)
\big(2 M m_3 \big(-m_1^4 + m_{12}^2 m_2^2 + m_{13}^2 m_2^2 - m_2^4 + m_{13}^2 m_{23}^2 \nonumber  \\
& + m_1^2 \left(m_{12}^2 - m_2^2 + m_{23}^2\right) - \left(m_{12}^2 + m_{13}^2 + m_{23}^2\right) m_3^2 + m_3^4\big)\big) \nonumber  \\
& + \operatorname{Re}\left(C_5 C_7^* + C_6 C_8^*\right)
\left(-2 m_1 m_2 \left(m_1^2 - m_{13}^2 + m_2^2 - m_{23}^2\right)\left(m_1^2 - m_{12}^2 - m_{13}^2 + m_2^2 - m_{23}^2 + m_3^2\right)\right). \nonumber 
\end{align}
As before, we don't present the interference between dimension 6 and 7 operators.  By replacing $m_{1,2,3}$ with $m_e$,  $m_\mu$, or $m_\nu$, one can get the finite mass results for 
\begin{gather*}
\tau \rightarrow \bar{\mu}\mu e, \text{ }
\tau \rightarrow \bar{e}e\mu, \text{ }\ell \to \ell' \nu \bar{\nu}.
\end{gather*}

\section*{Appendix C: Summary of EFT Operators}
In this section, we summarize the EFT operators for our decays.  This follows the approach in \cite{Chang:2022crb,Bradshaw:2023wco,Arzate:2023swz} and is a reanalysis of the four fermion operators in \cite{Bradshaw:2023wco}. In those works, the independent four body operators are determined at each mass dimension by looking at onshell two to two scattering processes.  In this paper, we need the four fermion operators for the case where two of the fermions might be indistinguishable.  Thus, we list all primary EFT operators for the general four fermion interactions involving  $F f' \bar{f} \bar{f}$ and $F f' \bar{f}'' \bar{f}'''$. These operators break up into primary operators and descendant operators, where descendants are primary operators dressed with contracted derivatives.  In terms of the on-shell amplitudes, the descendant amplitudes are merely Mandelstam invariants multiplying the primary amplitudes.  A useful cross check is that the number of independent operators at each dimension can be counted by the Hilbert Series approach \cite{PhysRevD.91.105014, Henning_2015, Lehman_2016, Henning:2015alf, Henning_2017, Gr_f_2021, Gr_f_2023}.  This calculation gives a generating function which tells how many operators are at each dimension (up to potential cancellations due to redundancies for some descendant operators, see \cite{Bradshaw:2023wco, Arzate:2023swz} for  detailed discussions).   The Hilbert series calculations give 
\begin{equation}
H_{F f' \bar{f} \bar{f}}= \frac{6q^6+2q^7+4q^8+6q^9-2q^{10}}{(1-q^2)(1-q^4)},
\end{equation}
\begin{equation}
H_{F f' \bar{f}'' \bar{f}'''} = \frac{10q^6+8q^7-2q^8}{(1-q^2)^2}.
\end{equation}
To use these, one Taylor expands in $q$ and a term $c q^d \in H$ naively says that there are $c$ independent operators at dimension $d$.  The caveat to this is that there can be descendant operators which are redundant.  Ignoring this caveat for now, the numerators count the number of primary operators and expanding the denominator gives the Mandelstam invariants.  For $F f' \bar{f}'' \bar{f}'''$, the two factors of $(1-q^2)$ is giving powers of the  independent Mandelstams $s$ and $t$, which are both mass dimension 2.  While for $F f' \bar{f} \bar{f}$, the denominator of $(1-q^2)(1-q^4)$ generates arbitrary Mandelstam factors  of $s$ and $(t-u)^2$,  which are respectively dimension 2 and 4,  Moreover, since $t\leftrightarrow u$ when exchanging the identical $\bar{f}$, $(t-u)^2$ is invariant under exchange of those identical particles.  In the numerators there are negative coefficients at dimension 10 and 8 respectively for the two cases.  These are correction terms that say that 2 descendants at those dimensions become redundant.  

The primary operators for $Ff'\bar{f}\bar{f}$ are listed in Table~\ref{tab:allamps} (left).  Consistent with the positive coefficients in the Hilbert series numerator, there are six operators at dimension six, two at dimension seven, four at dimension eight, and six at dimension nine.  The $-2q^{10}$ is due to the fact that at dimension 10, $s \mathcal{O}_{9}$ and $s\mathcal{O}_{12}$ become redundant, where $s$ is the Mandelstam between the two identical fermions.  That means one only needs to include descendants of $\mathcal{O}_{9}$ and $s\mathcal{O}_{12}$ with arbitrary factors of $(t-u)^2.$  

The primary operators for $F f' \bar{f}'' \bar{f}'''$ are listed in Table~\ref{tab:allamps} (right). Consistent with the Hilbert series numerator, there are 10 primary operators at dimension six and eight at dimension seven.  The $-2q^{8}$ agrees with our finding that  at dimension 8, $s \mathcal{O}_{9}$ and $s\mathcal{O}_{10}$ become redundant, where $s=(p_F+p_{f'})^2$.  Thus one only needs to use descendants of $\mathcal{O}_{9}$ and $s\mathcal{O}_{10}$ with arbitrary factors of $t=(p_F-p_{\bar{f}''})^2.$ 

\begin{table}[b]
\begin{center}
\centering
\renewcommand{\arraystretch}{1.25}
\tabcolsep6pt\begin{tabular}{|c|c|c|c|c|}
\hline
{$i$} & {$\mathcal{O}_i^{F f' \bar{f} \bar{f}}$}  &  {$d_{\mathcal{O}_i}$}  \\
\hline
1 &  $(\bar{f} P_R f') (\bar{f} P_R F)
$ &    \multirow{6}{*}{6} \\
2 &  $(\bar{f} P_L f') (\bar{f} P_L F)$ &      \\
3 &  $(\bar{f} \gamma^\alpha P_L f') (\bar{f} \gamma_\alpha P_L F)$ &    \\
4 &  $(\bar{f} \gamma^\alpha P_R f') (\bar{f} \gamma_\alpha P_L F) $ &       \\
5 &  $(\bar{f} \gamma^\alpha P_L f') (\bar{f} \gamma_\alpha P_R F) $ &         \\
6 &  $(\bar{f} \gamma^\alpha P_R f') (\bar{f} \gamma_\alpha P_R F) $ &         \\
\hline 
7 &  $
([\partial_\alpha\bar{f}]  P_L  f') (\bar{f} \gamma^\alpha P_L F)$ &    \multirow{2}{*}{7}  \\
8 &  $([\partial_\alpha\bar{f}]  P_R  f') (\bar{f} \gamma^\alpha P_R F)$  &   \\
\hline
9 &  $ ([{\overset\leftrightarrow{D}}\phantom{}^f_\alpha \bar{f}] P_R D^\alpha  f') (\bar{f} P_R F)
 $ &   \multirow{4}{*}{8}   \\
10 &  $ ([{\overset\leftrightarrow{D}}\phantom{}^f_\alpha \bar{f}] P_L D^\alpha  f') (\bar{f} P_R F)  $ &   \\
11 &  $ ([{\overset\leftrightarrow{D}}\phantom{}^f_\alpha \bar{f}] P_R D^\alpha  f') (\bar{f} P_L F) $ &    \\
12 &  $([{\overset\leftrightarrow{D}}\phantom{}^f_\alpha \bar{f}] P_L D^\alpha  f') (\bar{f} P_L F)$   &   \\
\hline
13 &  $([{\overset\leftrightarrow{D}}\phantom{}^f_\beta D_\alpha \bar{f}] P_R D^\beta  f') (\bar{f} \gamma^\alpha P_L F)$ &    \multirow{6}{*}{9}    \\
14 &  $([{\overset\leftrightarrow{D}}\phantom{}^f_\beta D_\alpha \bar{f}] P_L D^\beta  f') (\bar{f} \gamma^\alpha P_L F)$ &  \\
15 &  $([{\overset\leftrightarrow{D}}\phantom{}^f_\beta D_\alpha \bar{f}] P_R D^\beta  f') (\bar{f} \gamma^\alpha P_R F)$  &    \\
16 &  $([{\overset\leftrightarrow{D}}\phantom{}^f_\beta D_\alpha \bar{f}] P_L D^\beta  f') (\bar{f} \gamma^\alpha P_R F)$   &   \\

17 &  $([{\overset\leftrightarrow{D}}\phantom{}^f_\beta  \bar{f}] \gamma^\alpha P_L D^\beta  f') ([D_\alpha \bar{f}]  P_R F)$  &  \\

18 &  $([{\overset\leftrightarrow{D}}\phantom{}^f_\beta  \bar{f}] \gamma^\alpha P_R D^\beta  f') ([D_\alpha \bar{f}]  P_L F)$  & \\
\hline
\end{tabular}
\quad \quad \quad \quad \quad
\begin{tabular}{|c|c|c|c|c|}
\hline
 {$i$} &  {$\tilde{\mathcal{O}}_i^{F f' \bar{f}'' \bar{f}'''} $}  &  $d_{\mathcal{O}_i}$   \\
\hline
1 &  $(\bar{f}'' P_R f') (\bar{f}''' P_R F) $ &    \multirow{10}{*}{6}    \\
2  & $(\bar{f}'' P_L f') (\bar{f}''' P_R F) $ &   \\
3 &  $(\bar{f}'' P_R f') (\bar{f}''' P_L F) $ &   \\
4 & $(\bar{f}'' P_L f') (\bar{f}''' P_L F) $  &     \\
5 &  $(\bar{f}'' \gamma_\alpha P_L f') (\bar{f}''' \gamma^\alpha P_L F) $ &   \\
6 &   $(\bar{f}'' \gamma_\alpha P_R f') (\bar{f}''' \gamma^\alpha P_L F) $ &      \\
7 &   $(\bar{f}'' \gamma_\alpha P_L f') (\bar{f}''' \gamma^\alpha P_R F) $ &        \\
8 &   $(\bar{f}'' \gamma_\alpha P_R f') (\bar{f}''' \gamma^\alpha P_R F) $ &       \\
9 &   $(\bar{f}'' P_R \sigma_{\alpha\beta} f') (\bar{f}''' P_R \sigma^{\alpha\beta}  F) $ &     \\
10 &  $(\bar{f}'' P_L \sigma_{\alpha\beta} f') (\bar{f}''' P_L \sigma^{\alpha\beta}  F) $ &     \\
  \hline
11 &  $(\bar{f}'' P_R \partial_{\alpha} f') (\bar{f}''' \gamma^\alpha P_L   F) $ &    \multirow{8}{*}{7}  \\
12 &  $(\bar{f}'' P_L \partial_{\alpha} f') (\bar{f}''' \gamma^\alpha P_L   F) $ &   \\
13 &  $(\bar{f}'' P_R \partial_{\alpha} f') (\bar{f}''' \gamma^\alpha P_R   F) $  &  \\
14 &  $(\bar{f}'' P_L \partial_{\alpha} f') (\bar{f}''' \gamma^\alpha P_R   F) $  &    \\
15 &  $(\bar{f}''  \gamma^{\alpha} P_L f') (\bar{f}''' P_R \partial_\alpha    F) $ &    \\

16 &  $(\bar{f}''  \gamma^{\alpha} P_L f') (\bar{f}''' P_L \partial_\alpha    F) $  &    \\
17 &  $(\bar{f}''  \gamma^{\alpha} P_R f') (\bar{f}''' P_R \partial_\alpha    F) $ &  \\
18 &  $(\bar{f}''  \gamma^{\alpha} P_R f') (\bar{f}''' P_L \partial_\alpha    F) $  & \\
\hline
\end{tabular}

\caption{\label{tab:allamps}  \footnotesize Primary operators for $F f' \bar{f} \bar{f}$ (left) and $F f'\bar{f}'' \bar{f}'''$ (right).   For $F f' \bar{f} \bar{f}$, we've defined a back-forth derivative $\overset\leftrightarrow{D}\vphantom{D}^{f}_\alpha$, which acts only on the $\bar{f}$ fields.  For descendant operators, one adds contracted derivatives to get arbitrary Mandelstam factors that respect the exchange symmetry, i.e.~$s, (t-u)^2$.  At dimension 10, $s \mathcal{O}_{9}$ and $s\mathcal{O}_{12}$ become redundant.  Thus  one only needs to consider $\mathcal{O}_{9,12}$ descendants with arbitrary factors of $(t-u)^2$. For $F f' \bar{f}''\bar{f}'''$ descendant operators, one adds contracted derivatives to get arbitrary Mandelstam factors of $s, t$.  At dimension 8, $s \tilde{\mathcal{O}}_{9}$ and $s\tilde{\mathcal{O}}_{10}$ become redundant.  Thus  one only needs to consider $\tilde{\mathcal{O}}_{9,10}$ descendants with arbitrary factors of $t$. }
\end{center}
\end{table}

\bibliographystyle{utphys}
\bibliography{citations}

\end{document}